\documentclass[10pt, conference, compsocconf]{IEEEtran}
\IEEEoverridecommandlockouts
\usepackage{cite}
\usepackage{amsmath,amssymb,amsfonts}
\usepackage{algorithmic}
\usepackage{graphicx}
\usepackage{textcomp}
\usepackage{xcolor}

\usepackage{booktabs}
\usepackage{multirow}
\usepackage{url}

\usepackage{bm}

\usepackage{enumitem}
\usepackage{subfigure}
\usepackage{amsthm}

\usepackage{caption}
\newcommand{\QZY}{\textcolor{black}}

\def\BibTeX{{\rm B\kern-.05em{\sc i\kern-.025em b}\kern-.08em
    T\kern-.1667em\lower.7ex\hbox{E}\kern-.125emX}}

\newtheorem{definition}{Definition}

\begin{document}

\title{\LARGE \textbf{Tree Structure-Aware Graph Representation Learning via Integrated
Hierarchical Aggregation and Relational Metric Learning}
}
% Pengyang
%\title{\LARGE \textbf{
%Tree Structure-Aware Graph Representation Learning via Integrated
%Hierarchical Aggregation and Relational Metric Learning
%}}

\author{\IEEEauthorblockN{Ziyue Qiao$^{1,2}$, Pengyang Wang$^3$, Yanjie Fu$^3$,  Yi Du$^{1,*}$\thanks{$^*$Corresponding author.},  Pengfei Wang$^4$, Yuanchun Zhou$^1$}
\IEEEauthorblockA{\textit{$^1$Computer Network Information Center, Chinese Academy of Sciences, Beijing} \\
\textit{$^2$University of Chinese Academy of Sciences, Beijing} \\
\textit{$^3$Department of Computer Science, University of Central Florida, Orlando}\\
\textit{$^4$Alibaba DAMO Academy, Alibaba Group, China}\\
qiaoziyue@cnic.cn, \{pengyang.wang@knights., yanjie.fu@\}ucf.edu, \{duyi, zyc\}@cnic.cn}
\thanks{The research is supported by the Natural Science Foundation of China under Grant No. 61836013, the National Key Research and Development Plan of China (No. 2016YFB0501901), Ministry of Science and Technology Innovation Methods Special work Project under grant 2019IM020100,  Beijing Nova Program of Science and Technology under Grant No. Z191100001119090.}
}

\maketitle

\begin{abstract}

While Graph Neural Network (GNN) has shown superiority in learning node representations of homogeneous graphs, leveraging GNN on heterogeneous graphs remains a challenging problem.
The dominating reason is that GNN learns node representations by aggregating neighbors' information regardless of node types.
Some work is proposed to alleviate such issue by exploiting relations or meta-path to sample neighbors with distinct categories, then use attention mechanism to learn different importance for different categories.
However, one limitation is that the learned  representations  for different  types  of  nodes  should  own  different  feature  spaces, while  all  the  above  work  still  project  node  representations into one feature space.
Moreover, after exploring massive heterogeneous graphs, we identify a fact that multiple nodes with the same type always connect to a node with another type, which reveals the many-to-one schema, {\it a.k.a.} the hierarchical tree structure.
But all the above work cannot preserve such tree structure, since the exact multi-hop path correlation from  neighbors  to  the  target  node  would  be  erased  through aggregation.
Therefore, to overcome the limitations of the literature, we propose T-GNN, a tree structure-aware graph neural network model for graph representation learning.
Specifically, the proposed T-GNN consists of two modules: (1) the integrated  hierarchical aggregation module and (2) the relational metric learning module.
The integrated  hierarchical aggregation module aims to preserve the tree structure by combining GNN  with  gated  recurrent  unit  to integrate the hierarchical and sequential neighborhood information on the tree structure to node representations.
The relational metric learning module aims to preserve the heterogeneity by embedding  each  type  of  nodes  into  a type-specific space with distinct distribution based on similarity metrics.
In this way, our proposed T-GNN is capable of simultaneously preserving the heterogeneity and the tree structure inherent in heterogeneous graphs.
Finally, we conduct extensive experiments to show the outstanding performance of T-GNN in tasks of node clustering and classification, inductive node clustering and classification, and link prediction.

\end{abstract}


\IEEEpeerreviewmaketitle

\section{INTRODUCTION}
Graph Neural Network (GNN) is a family of graph representation learning approaches that encode node features into low-dimensional representation vectors by aggregating neighbors' information \cite{zhou2018graph}.
GNN has drawn increasing attention in recent decades, due to the superior performance in tremendous real-world applications, such as recommender systems~\cite{fan2019metapath, wu2020joint}, urban computing~\cite{wang2019adversarial, wang2020exploiting}, chemistry~\cite{coley2019graph}, etc.

\begin{figure}[!t]
\setlength{\belowcaptionskip}{-0.3cm}
  \centering
  \includegraphics[width=0.48\textwidth]{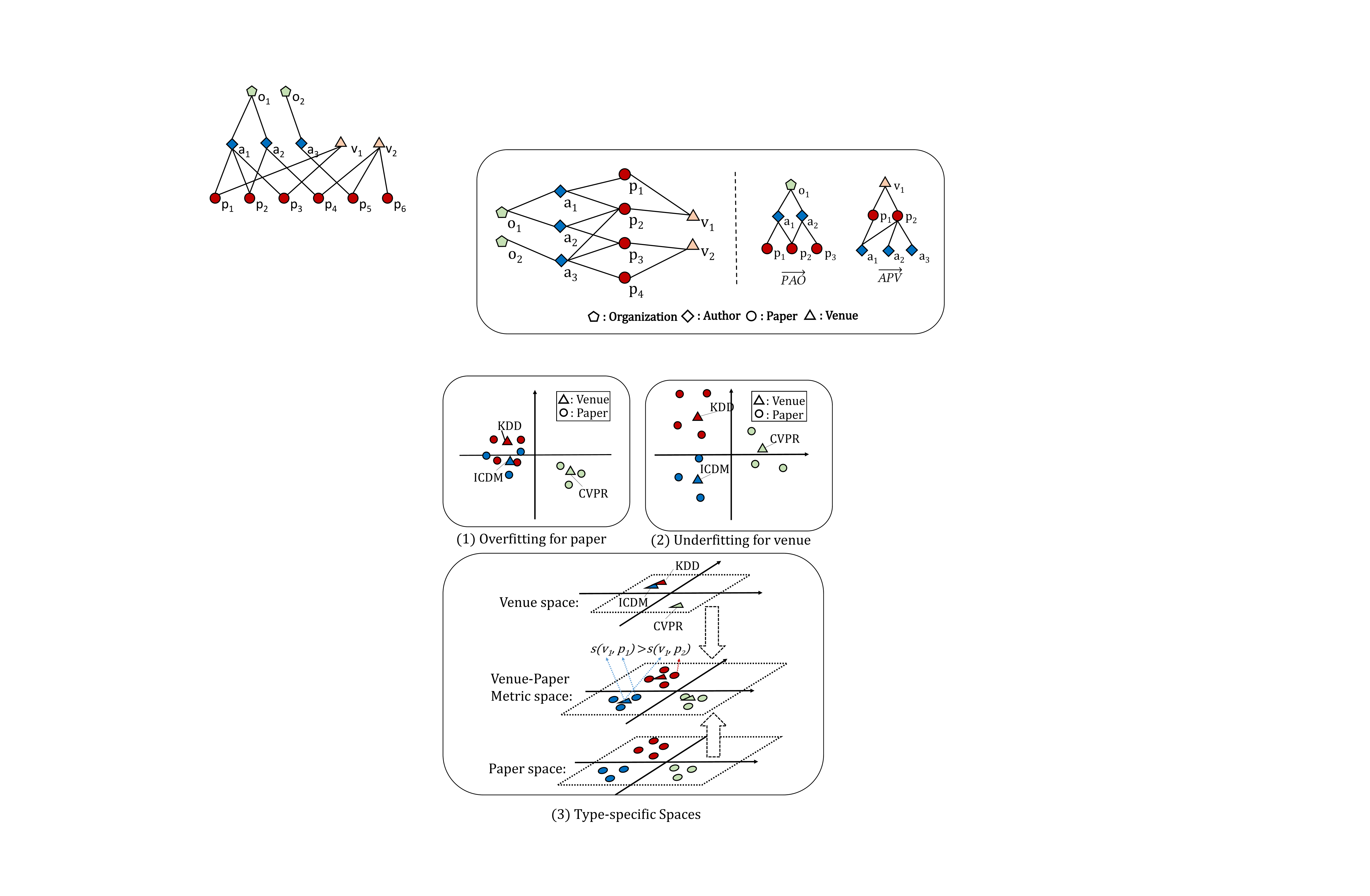}
  \caption{An illustration of hierarchical tree structures in academic heterogeneous graph.}
  \label{HNet}
\end{figure}

While GNN models conform well with the learning tasks on homogeneous graphs, leveraging GNN on learning node representations for heterogeneous graphs remains a challenging problem.
First of all, heterogeneous graphs consist of nodes with different types, but vanilla GNN indiscriminately aggregates neighbors' information regardless of node types.
Second, the relations between different types of nodes usually reveal the structure of many-to-one, {\it a.k.a.} the hierarchical tree structure, which indicates the schema that multiple nodes with the same type connect to a node with another type. For example, Figure \ref{HNet} shows an instance in the academia graph, papers $p_1$, $p_2$ and $p_3$ are linked to author $a_1$ and $a_2$ by the relation Paper$\xrightarrow{authored}$ Author (abbreviated as $\overrightarrow{PA}$), also authors $a_1$ and $a_2$ are linked to organization $o_1$ by the relation Author$\xrightarrow{employed}$ Organization (abbreviated as $\overrightarrow{AO}$).
By composing $\overrightarrow{PA}$ and $\overrightarrow{AO}$, the nodes of type P, A and O form a three-level hierarchical tree with structure $\overrightarrow{PAO}$ and $o_1$ as the root node, this tree implies a two-hop structured neighborhood for $o_1$, where the papers are in the ground level and the authors are in the second level.

\iffalse
In literature, some work is proposed to alleviate such issue, which can be categorized into two groups, (1) meta-path based methods, which extracts meta paths to preserve the heterogeneity~\cite{sankar2019meta,fan2019metapath};
and (2) attention-based methods, which assigns different attentions to different node types as aggregating neighbors~\cite{wang2019heterogeneous,zhang2019heterogeneous}.
However, there still exists limitations.
First, the learned representations for different types of nodes should own different feature spaces, while all the above work still project node representations into one feature space.
Second, for meta-path based methods, given the target node, meta-paths directly transfer the information between the two end nodes of the meta paths, while ignores the relay nodes in the middle.
Third, for the attention based methods, due to the nature of aggregation, although GNN can propagate information of multi-hop neighborhoods by stacking multiple layers, the exact multi-hop path correlation from neighbors to the target node would be erased through aggregation.
Consequently, the tree structure would be ignored.
\fi

In literature, some work are proposed to alleviate such issue.
The main idea is sampling neighbors with distinct categories via different types of relations or meta-path (i,e,. composed relations), and use attention mechanism or coefficients to assign different importance to different categories of nodes as aggregating neighbors~\cite{sankar2019meta, fu2020magnn, wang2019heterogeneous, qiao2019unsupervised, zhang2019heterogeneous}.
% , then the encoded node representations are learned by various optimization models.
However, there still exists limitations.
First, some methods\cite{schlichtkrull2018modeling, zhu2019relation, zhang2019heterogeneous, wang2020generic} only differentiate neighbor information by atomic relations, i.e., the one-hop links, and propagate information of multi-hop neighborhoods by stacking multiple layers. However, the exact multi-hop path correlation from neighbors to the target node would be erased through aggregation.
Second, for some methods\cite{wang2019heterogeneous, fan2019metapath} using meta-path to sample neighbors in multi-hop, given the target node, meta-paths directly transfer the information between the two end nodes of the meta paths, while ignore the relay nodes in the middle. On the other hand, meta-path based neighborhoods preserve less structural information relatively than the tree-based neighborhoods.
Third, the learned representations for different types of nodes should own different feature spaces, while all the above work still projects node representations into one feature space.

Therefore, to overcome the limitations of the literature, we propose a \textbf{T}ree structure-aware \textbf{G}raph \textbf{N}eural \textbf{N}etwork(T-GNN) for heterogeneous graph representation learning. Next, we outline the key steps and insights of our proposed T-GNN.

\begin{figure}[!t]
\setlength{\belowcaptionskip}{-0.3cm}
  \centering
  \includegraphics[width=0.35\textwidth]{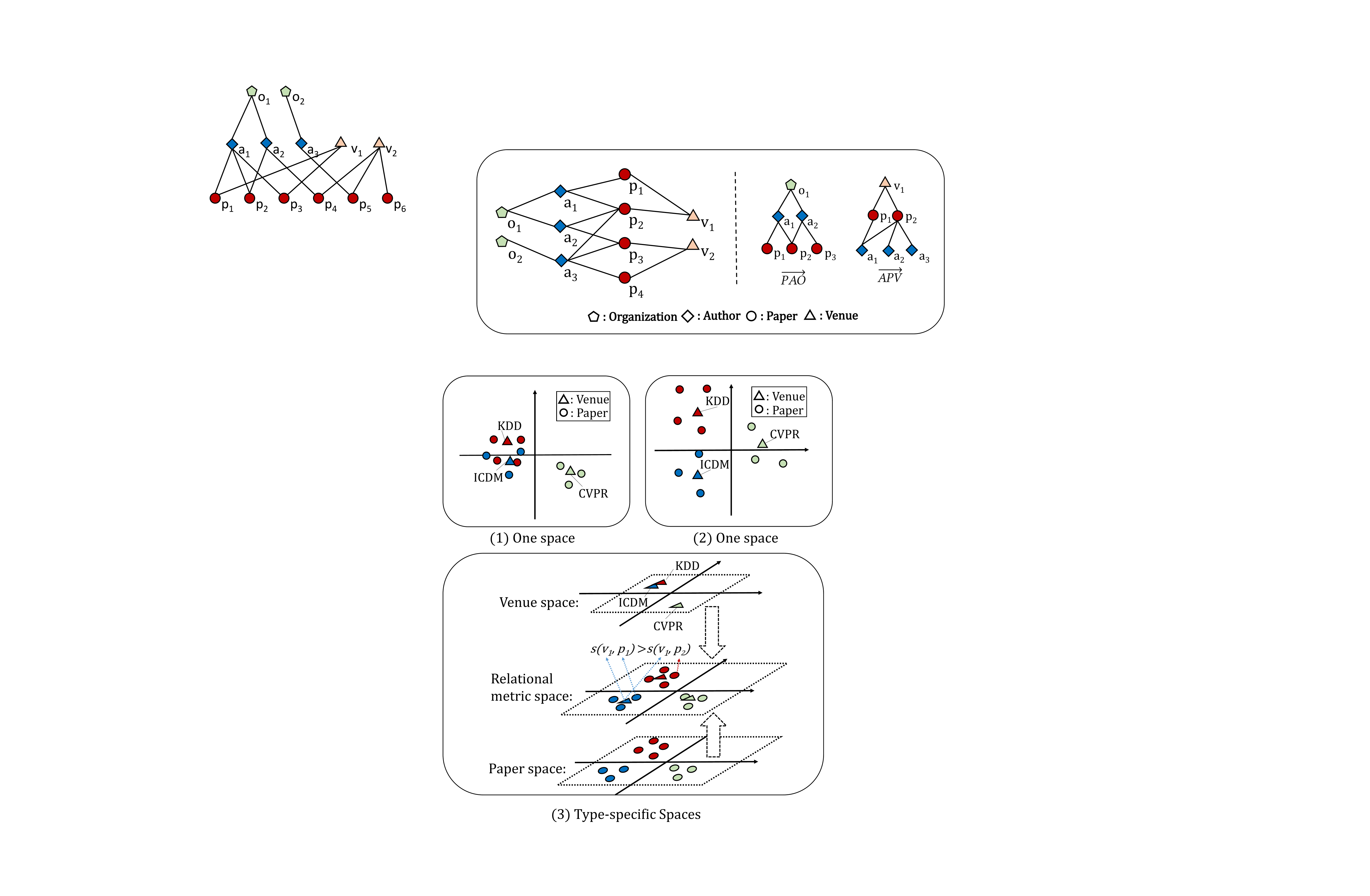}
  \caption{An illustration of respectively embedding three venues and their corresponding papers into same spaces and type-specific spaces. (1)(2) shows if different types of nodes are embeded into one space. (3) shows by embedding nodes into type-specific spaces.}
  \label{distribute}
\end{figure}

{\bf Preserving Tree Structure via Integrated Hierarchical Aggregation.}
As mentioned above, multi-hop neighborhoods in heterogeneous graphs can be decomposed to different trees, which can denote different organizational forms of neighbors and relations.
Also on these trees, the neighbor information with different distance can be further formulated as a sequence with variable length from the bottom level to the top level.
This indicates that we can sequentially aggregate the information of lower level nodes and propagate it to the higher level nodes to finally contribute to the root nodes' representations.
Therefore, we propose an integrated hierarchical aggregation module to preserve tree structures of heterogeneous graphs.
Specifically, we  differentiate  tree  structured  neighborhoods by introducing hierarchical tree schema, and partition neighborhoods into organized trees with different categories. Then we combine GNN models with Gated Recurrent Unit (GRU) modules to aggregate hierarchical and sequential neighborhood information on these trees to node representations.

{\bf Preserving Heterogeneity via Relational Metric Learning.}
Most heterogeneous graph representation learning methods aim to embed different types of nodes into the same feature space. However, in one space, different types of nodes share the same polynomial distribution but the similarity between nodes of different types could be incompatible and dampen with each other.

\iffalse
For example, in Figure \ref{distribute}(a), each venue node publishes multiple papers.
If the distance between two similar venues are prefer to get close, the papers published by these two venues would tend to be inseparable.
An ideal feature distribution for papers is that each paper is close to its own venue, and papers from different venue would be relatively distant from each other.
%\PW{But in this case, venue nodes need larger scale of distribution between each other than papers as shown in Figure \ref{distribute}(b), this may lead to the representation learning of venues does not converge better than that of papers.: ???}
But in this case, similar venues are distant with each other, this may lead to the proximity between venues does not preserve better than that of papers.
\fi

For example in Figure \ref{distribute}(a),
the closeness of two similar venues would results in the papers published by these two venues tend to be inseparable, while in Figure \ref{distribute}(b), well separated papers makes their two venues distant and the proximity of them would not be well preserved.
Such incompatibility may also exists between other types of nodes in one embedding space. Model optimization usually need to learn specific distributions of nodes for certain tasks.
Therefore, we propose a relational metric learning module to embed each type of nodes into a type-specific space with independent distribution.
Specifically, we introduce metrics to measure the similarities between nodes with different types, as shown in Figure \ref{distribute}(c), the similarity between nodes with same type is calculated in their type space, and that between nodes with different types is calculated in a relation type-specific metric space.
Finally, we leverage a relation-aware graph context loss based on random walk and negative sampling to train the whole model.

% As shown in the example in Figure \ref{distribute}(c), the similarity between venues and papers can be calculated in a independent metric space, the closeness between venue in their own space could not effect papers' distributions on paper space and vice versa.
%Specifically, we introduce two kinds of metrics respectively  to  measure  the  similarities  between nodes  with  different  types,  and  leverage a relation-aware graph context loss based  on random walks and negative sampling to train the  model.
%\PW{it may need more details.}

In summary, we  propose a T-GNN model to improve the capability of GNN on learning node representations of heterogeneous graphs.
The proposed T-GNN includes two key modules, (1) hierarchical aggregation module, which aims to preserve the tree structure by integrating GNN with GRU;
(2) relational metric learning module, which aims to preserve the heterogeneity by mapping different type of nodes into different embedding space with incorporating similarity measurement.
The main contributions of our work are summarized as follow.

\begin{enumerate}
\item To our best knowledge, we are the first to introduce the tree structures inherent in heterongenous graphs into the architecture of graph neural network model to learn node representations.
\item We propose a tree structure-aware graph neural network model named T-GNN, which contains aggregation modules and sequence propagation modules to capture both the node attributes and the information in multi-hop neighborhoods with different hierarchical organizations into node representations.
\item We propose a relational metric learning module for unsupervised graph representation learning that embed nodes into type-specific spaces with independent distributions.
\item We conduct extensive experiments to evaluate the performance of our proposed model on three datasets. The results demonstrate the superiority of our proposed method over several state-of-the-art methods.

\end{enumerate}

\section{PRELIMINARIES}
In this section, we give formal definitions of some related concepts and notations.

\begin{definition}[\textbf{Heterogeneous Graph}]
A heterogeneous graph is denoted as $G=(N,E,\mathcal{T},\mathcal{R})$, in which each node $n\in{N}$ is associated with a node type mapping functions $\phi(n):N\to\mathcal{T}$ and each edge $e\in{E}$ is associated with a relation type mapping function $\phi(e):E\to{\mathcal{R}}$. This suggests that G has multiple node types and relation types, and $|\mathcal{T}|+|\mathcal{R}|>2$.
\end{definition}

\begin{figure}
\setlength{\belowcaptionskip}{-0.3cm}
  \centering
  \includegraphics[width=0.45\textwidth]{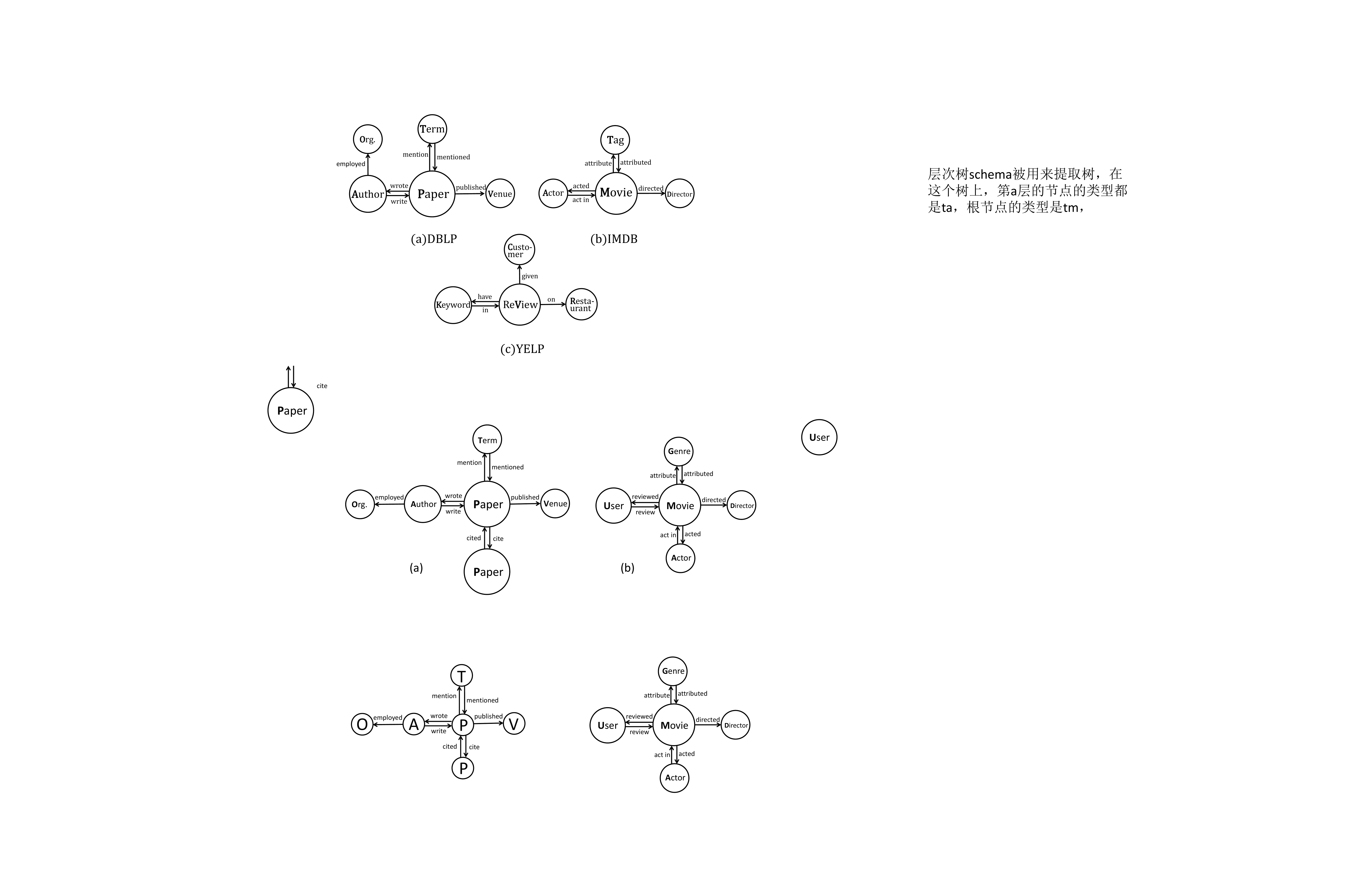}
  \caption{Schemas of three heterogeneous graphs. (a) DBLP Academic graph. (b) IMDB Movie graph. (c) YELP review graph. Each solid line arrows represent a relation type, each circles represent one node type and the bold letters represent their abbreviations.}
  \label{imgHR}
\end{figure}

Figure \ref{imgHR} shows the schemas of three heterogeneous graphs. We resolve the undirected links between nodes into fine-grained directed relations to present the hierarchical structures. The directions of relations are consistent with the structures of many-to-one between different types of nodes.

%\PW{What is the purpose of this paragraph? - As GNNs are information aggregating models usually utilized on subgraphs with the structure of many neighbors with same type connecting to the target node and sharing the same transformer, we can resolve all the undirected links between nodes in heterogeneous graph into fine-grained directed relations to guide the aggregating direction between different types of nodes. We assign the links connecting each two types of nodes with the structure of many-to-one to a distinct relation type. Specifically, if more than one nodes with the same type $t_a \in \mathcal{T}$ connect to a node with type $t_b \in \mathcal{T}$, we assume there are directed edges between from $t_a$ to $t_b$ and these edges have a distinct relation type $r\in \mathcal{R}$. For example, the relation $Paper\xrightarrow{published}Venue$ has the structure that many papers connect to one venue. There could be two reverse relations between two types of nodes, such as $Paper\xrightarrow{writed}Author$ and $Author\xrightarrow{write}Paper$. We find this is applicable for extracting different kinds of structured heterogeneous graphs, the examples are shown in Figure \ref{imgHR}, the directed line arrows represent the relations and their direction.}

\begin{definition}[\textbf{Meta-path}]
%\PW{It would be better provide some examples.}
In a heterogeneous graph $G=(N,E,\mathcal{T},\mathcal{R})$, a meta-path is defined as a path in the form of $t_0\xrightarrow{r_1}t_1\xrightarrow{r_2}...\xrightarrow{r_{l}}t_{l}$ (abbreviated as $t_0t_1...t_{l}$), where $r_i\in\mathcal{R}$ and $t_i\in\mathcal{T}$, the sequence of relations represents the composite relation $r = r_1\circ r_2\circ ...\circ r_{l}$ between the node type $t_0$ and $t_l$.
\end{definition}

For example, we can design $PAP$, $APV$ as meta-paths, $PAP$ means two papers are coauthored by a same person, $APV$ means an author publish a paper on a venue. Meta-paths are widely utilized to extract paths in previous works. Similarly, in our paper, we define hierarchical tree schemas to extract hierarchical trees.
%Meta-paths, as the name implies, are usually defined to extract paths in previous works, To make a difference, we define hierarchical tree schema to extract hierarchical trees.

\begin{definition}[\textbf{Hierarchical Tree Schema}]
A Hierarchical Tree Schema $s$ is a directed meta-path in the form of $s = \{t_0\xrightarrow{r_1}t_1\xrightarrow{r_2}...\xrightarrow{r_{m}}t_{m}\}$ (abbreviated as $\overrightarrow{t_0 t_1...t_{m}}$), where the direction of each $r_i$ is from $t_{i-1}$ to $t_i$. These schemas are used to extract trees where $t_m$ is the type of root node and each node in the $i$th level has the type $t_i$, representing the $(m-i)$-hop neighbors of the root node.
\end{definition}

\begin{definition}[\textbf{Hierarchical Tree Structured Neighborhood}]
Given a hierarchical tree schema $s = \{t_0\xrightarrow{r_1}t_1\xrightarrow{r_2}...\xrightarrow{r_{m}}t_{m}\}$ and a node $n_o$ with type $t_{m}$, we first sample the neighbors of $n_o$ connecting by relation $r_m$ to the $(m-1)$th level, then we sample neighbors of nodes in the $(m-1)$th level connecting by relation $r_{m-1}$ to the $(m-2)$th level, an so on, sample nodes until to the $0$th level.
Finally, we can obtain a hierarchical tree with $n_o$ in the level $m$ as the root node, and the nodes in level $m-k$ with type $t_{m-k}$ is $n_o$'s $k$-hop neighbors.
\end{definition}

For example, in academic dataset, the hierarchical tree schema $\overrightarrow{TPV}$ represent a kind of structured neighborhood for venue nodes, where venue is the root nodes on the 2-th level, its papers are on the 1-th level, and the terms of these papers are on the 0-th level. $\overrightarrow{AMD}$ in movie dataset represents a structured neighborhood for each director, containing its movies on the 1-th level and the actors of these movies on the 0-th level.

Given a node $n_o$ with node type $t_m=\phi(n_o)$, we can extract $k$ kinds of hierarchical tree schemas ended with $n_o$'s type: $\mathcal{S}_t=\{s_1,s_2,...,s_k\}$, that is, $\forall s\in \mathcal{S}_t, s = \{t_0\xrightarrow{r_1}t_1\xrightarrow{r_2}...\xrightarrow{r_{m}}t_{m}\}$.
To differentiate the neighborhoods that different schemas refer to, we define each scheme in $\mathcal{S}_t$ is not a subsequence of another, i,e,. $\forall s_i, s_j \in \mathcal{S}_t \bigwedge i\neq j \rightarrow s_i \nsubseteq s_j$.

These hierarchical tree schemas partition $n_o$'s neighborhoods into different categories of hierarchical trees. These hierarchical trees preserve different semantics and sequence information and are discriminated in aggregation model. For example, $\overrightarrow{TPV}$ and $\overrightarrow{APV}$ represent two kinds of structured neighborhood of venues. For $\overrightarrow{TPV}$, first term information is aggregated to papers, and then papers which contain the terms' information are aggregated to the venues, while for $\overrightarrow{APV}$, the aggregation order is authors, papers and venues.

\begin{definition}[\textbf{Heterogeneous Graph Representation Learning}]
Given a heterogeneous graph $G=(N,E,\mathcal{T},\mathcal{R})$
learning the representation is to learn node type-specific mapping functions $\{f_{t}: N_{t}\to \mathbb{R}^{d'}_{t}\}, t \in \mathcal{T}$, which projects the nodes in $N_{t}$(node sets of type $t$) to a representation vector in the $d'$-dimensional latent space $\mathbb{R}^{d'}_{t}$, corresponding to the feature space of the node type $t$.
\end{definition}

%Distinct with traditional heterogeneous graph embedding methods which embed different types of nodes into one embedding space. We consider that nodes with different types should have different distributions, so we propose a heterogeneous graph embedding method that embed each type of nodes into a type-specific space that preserve the node features and network local neighborhoods  in node representations.

\begin{figure*}
\setlength{\belowcaptionskip}{-0.3cm}
  \centering
  \includegraphics[width=1\textwidth]{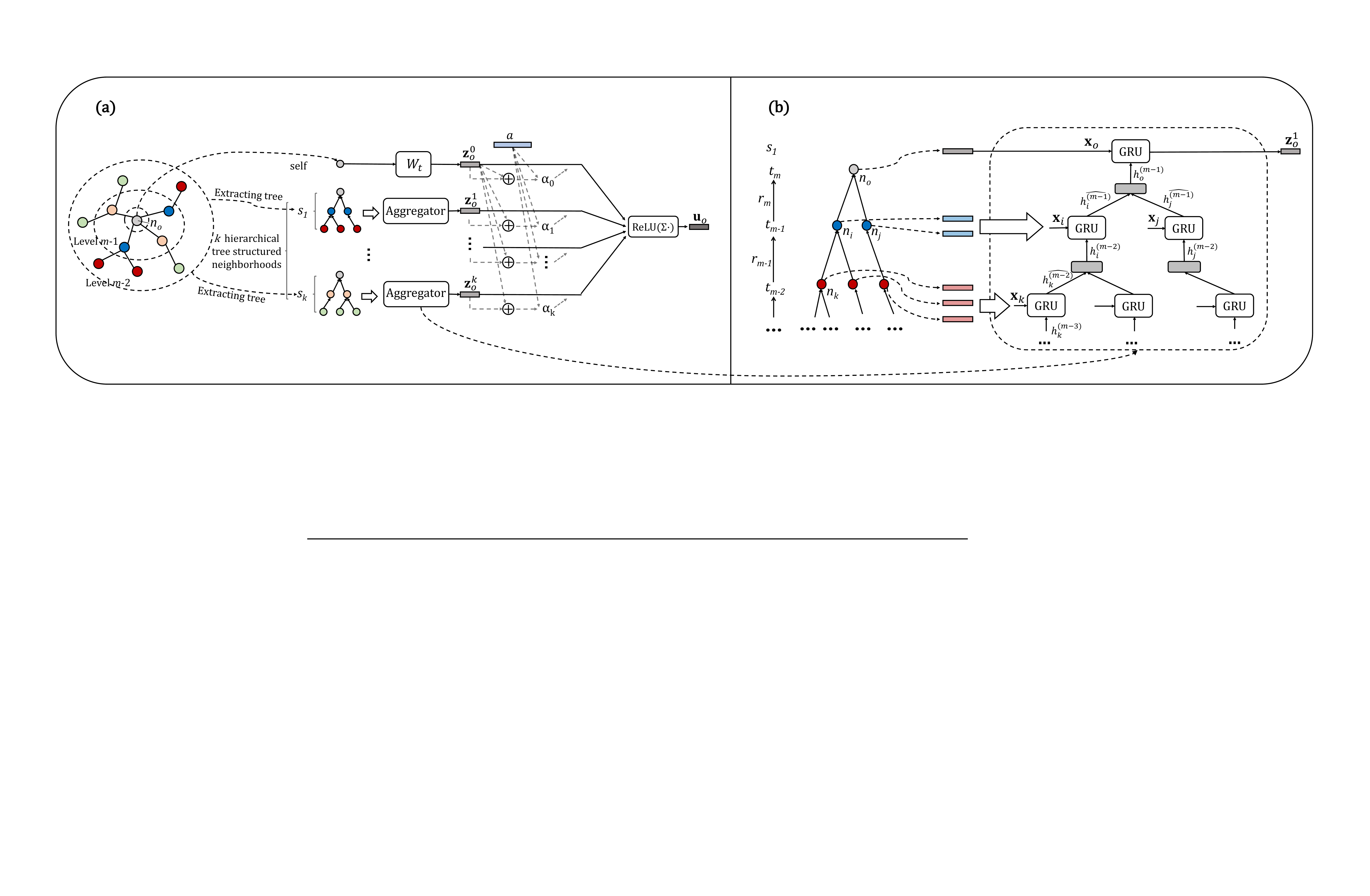}
  \caption{An illustration of our proposed propagation model. (a) The whole propagation model which consists of hierarchical tree structured neighborhood extracting, hierarchical aggregating and representation integrating. (b) The architecture of the proposed hierarchical aggregation.}
  \label{framework}
\end{figure*}

\section{PROPOSED MODEL}
In this section, we introduce the T-GNN model for heterogeneous graph representation learning.
Specifically, we combine the gated recurrent neural networks and GNN to process the information on hierarchical tree structured neighborhoods. Then we use attention mechanism to measure the importance of the information aggregated on different trees and integrate them into node representations. Finally, we present a relational metric learning module for model optimization.

\subsection{Hierarchical Aggregation}
For a node $n_o$, we have discussed we can use hierarchical tree schemas to divide its multi-hop neighborhood into multiple structured trees, each of those represents a distinctive neighborhood for $n_o$.
Given the hierarchical tree extracted by one schema $s$ where $n_o$ is the root node, and the initial feature vectors $\mathbf{x}_i \in \mathbb{R}^{d_t}$($d_t$: the initial feature vector dimension) of each node on this tree, our intuition is to aggregate the information from all the neighbors under $n_o$'s level on the tree as well as $\mathbf{x}_o$ to form its representations.

As the neighbor information in the propagation can be regarded as sequential inputs divided by levels and GNN can not process sequential data.
We conduct  \textbf{G}ated \textbf{R}ecurrent \textbf{U}nit (GRU) module combining with GNN on this tree to encode $\mathbf{x}_o$ and its neighbors information into a schema-specific output $\mathbf{z}^{s}_o$, the basic recurrence of the aggregation modules and propagation modules on hierarchical trees is:

\begin{equation}
\label{recurrence}
\begin{aligned}
&\widehat{h^{(0)}_i} = \mathbf{x}_i,\quad \phi (n_i) = t_0\\
&\widehat{h^{(a)}_i} = GRU(\mathbf{x}_i, h_i^{(a-1)}),\quad  \phi (n_i) = t_a\\
&h_i^{(a-1)} =  AGGREGATE_{r_a}(\{\widehat{h_j^{(a-1)}}, n_j\in N_i^{r_a}\})\\
\end{aligned}
\end{equation}

where $0<a\leq m$, $\widehat{h^{(a)}_i}$ represents the output hidden sate of node $n_i$ which is on the $a$th level of the tree with schema $s$, i,e, . The hidden state $h_i^{(a-1)}$ represents the neighborhood message of $n_i$ passing from all its neighbors on the $(a-1)$th level. The $GRU(\mathbf{x}_i, h_i^{(a-1)})$ is formulated as:

\begin{equation}
\label{GRU}
\begin{aligned}
&z_t = \sigma(A_z \mathbf{x}_i + B_z h_i^{(a-1)} )\\
&r_t = \sigma(A_r \mathbf{x}_i + B_r h_i^{(a-1)})\\
&\tilde{h}^{(a)}_i = tanh[A_h \mathbf{x}_i + B_h(r_t \circ h_i^{(a-1)})]\\
&\widehat{h^{(a)}_i} = z_t\circ h_i^{(a-1)} + (1- z_t)\circ\tilde{h}^{(a)}_i
\end{aligned}
\end{equation}

Where $\sigma(x)=\frac{1}{1+e^{-x}}$ is the sigmoid function and $\circ$ is element-wise multiplication.
%The aggregator function $AGGREGATE_{r_a}(\cdot)$ is \PW{what is relation type-specific? There are ``relation'', ``edge'', ``link'' in this paper, please unify} relation type-specific information aggregator of $n_i$'s sub-level neighbors,
\QZY{The aggregator function $AGGREGATE_{r_a}(\cdot)$ is a information aggregator of $n_i$'s sub-level neighbors specific for relation type $r_a$,}
which could be a mean pooling layer, max pooling layer, or a GNN aggregator, such as
mean aggregator, LSTM aggregator, pooling aggregator introduced in GraphSAGE\cite{hamilton2017inductive}, graph attentional layer introduced in GAT\cite{velivckovic2017graph}. In our paper, we use weighted mean aggregator, then the $h_i^{(a-1)}$ in Eq.\ref{recurrence} is reformulated as:

\begin{equation}
\label{GNN}
h_i^{(a-1)} = c^{r_a}_{ij}\sum_{n_j\in{N^{r_a}_i}}W_{r_a} \widehat{h^{(a-1)}_j}
\end{equation}

where $r_a$ is the relation type between level $a$ and level $a-1$ on $s$ that connect $n_i$ with its sub-level neighbors, $N^r_i$ denotes the set of neighbors indices of node $n_i$ under the relation $r\in\mathcal{R}$. $W_{r}$ is a trainable weight matrix specific for relation type $r$. $c^r_{ij}$ is the weight or normalization constant of the edge $(n_i,n_j)$  with type $r$.

After the $m$ levels' propagation on hierarchical trees structured neighborhood, the hidden output of node $n_o$ on schema $s$ can be obtained by:

\begin{equation}
\mathbf{z}_o = GRU(\mathbf{x}_o, h_o^{(m-1)})
\end{equation}

Where $\mathbf{z}_o \in \mathbb{R}^{d'}$($d'$: representation dimension). To make model tuning easy, we also set the dimension of hidden states as $d'$. In the neighbor information aggregation, parameters in GRU modules are shared and those in GNN modules are shared for same relation types. In practice, to save the calculation time, we use tensor operation and avoid the repeated aggregating on different trees with the same schemas, for example, for schema $\overrightarrow{TP}$ of paper nodes and $\overrightarrow{TPA}$ of author nodes, the aggregating processes from $T$ to $P$ are same for two schema and we only calculate once.

We have noticed that the gated neural units have been applied to some graph neural network models\cite{beck2018graph,seo2018structured,bresson2017residual}, majority of which are  originated from GGNN\cite{li2015gated}. %\PW{not clear - The main idea is that in the $k$-layer propagation of GNNs, the neighbor information of target nodes is aggregated by GNN in each layer, the $k$ hidden layer outputs are regarded as the sequenced information to be processed by GRUs.}
\QZY{Their main idea is stacking $k$-layer GNNs, and using GRU module to process the $k$ hidden layer outputs of each node as the sequenced information.}
Our proposed model is different from this idea: 1) We use the GRU module to help GNN directly aggregates the sequence information composed of neighbor nodes with arbitrary length,
%not the hidden layer outputs \PW{what do you mean by ``not the hidden layer outputs''};
2) Instead of stacking a fixed number of layers, our proposed T-GNN can stack different numbers of propagation layers for different node types according to their neighborhood depths.

\subsection{Neighborhood Information Integrating}
Then given the hierarchical tree schema set $\mathcal{S}_t=\{s_1,s_2,...,s_k\}$ of $n_o$, we can obtain the hidden output set $\mathbf{Z}_o = \{\mathbf{z}^{1}_o,\mathbf{z}^{2}_o,...,\mathbf{z}^{k}_o\}$ of $n_o$ and $\mathbf{z}^{i}_o\in \mathbb{R}^{d'}$. These hidden representations %\PW{You use ``embeddings'' and ``representations'', please unify}
that contain different neighborhood information may make different contributions to $n_o$'s final representation.
We employ the attention mechanism to combine these $k$ hidden representations with $n_o$'s feature vectors into the final representation vector of $n_o$, formulated as:

\begin{equation}
\mathbf{u}_o= ReLU\left(\alpha^o_o\cdot W_t \mathbf{x}_o + \sum_{\mathbf{z}^i_o\in\mathbf{Z}_o}\alpha^i_o\cdot \mathbf{z}^i_o \right)
\end{equation}

Where $\mathbf{u}_o\in \mathbb{R}^{d'}$ is the final representation vector of $n_o$, $\alpha^*_o$ indicates the importance of different hidden representations for $n_o$,
$W_t\in \mathbb{R}^{d'\times d}$ is a type-specific linear transformation to project the features vectors of nodes with type $t$ into a hidden representation vector, which represent the %self-information \PW{what is self-information?} of each node.
\QZY{feature information aggregated from each node itself}. For brevity, we denote that $\mathbf{z}^{0}_o = W_t \mathbf{x}_o$ and add it into representation set $\mathbf{Z}_o$, then $\mathbf{u}_o$ can be reformulated as:

\begin{equation}
\mathbf{u}_o= ReLU\left(\sum_{\mathbf{z}^{i}_o\in\mathbf{Z}_o}\alpha^{i}_o\cdot \mathbf{z}^{i}_o \right) \\
\end{equation}

We leverage self-attention to learn the attention weights $\alpha^{i}_o$, which are calculated via softmax function:

\begin{equation}
\alpha^{i}_o = \frac{exp\{LeakyReLU(a^T[\mathbf{z}^{0}_o||\mathbf{z}^{i}_o])\}}{\sum_{\mathbf{z}^{j}_o\in\mathbf{Z}_o}exp\{LeakyReLU(a^T[\mathbf{z}^{0}_o||\mathbf{z}^{j}_o])\}}
\end{equation}
which can be interpreted as the importance of the hidden representation $\mathbf{z}^{i}_o$, the higher the $\alpha^{i}_o$, the more important $\mathbf{z}^{i}_o$. $LeakyReLU$ denotes the leaky version of a Rectified Linear Unit, $a\in \mathbb{R}^{1\times 2d'}$ is the attention parameter.

Then we can apply the final node representation vector $\mathbf{u}_o$ to the loss functions in supervised tasks, semi-supervised tasks, or unsupervised tasks, and optimize the propagation model via back propagation.

\subsection{Optimization via Relational Metric learning}
\label{OM}

To perform heterogeneous graph representation learning, in this section, we leverage a graph context loss to optimize the model. Given a heterogeneous graph $G=(N,E,\mathcal{T},\mathcal{R})$, we aim to learn effective node representations by maximizing the probability of any node $n_i$ having its neighbor node $n_c$:

\begin{equation}
\label{objective}
\arg\max_\theta\sum_{n_i\in N}\sum_{r\in{\mathcal{R}}}\sum_{n_c\in N_r(n_i)}\log p(n_c|n_i,\theta)
\end{equation}
where $N_r(n_i)$ is a set of neighbors of $n_i$, and $\forall n_c\in N_r(n_i)$ connects $n_i$ with relation $r$, $\theta$ represents the parameters of the whole model. $p(n_c|n_i,\theta)$ is the probability of any node $n_i$ having its neighbor node $n_c$, defined as a softmax function:

\begin{equation}
\label{pro}
p(n_c|n_i,\theta) = \frac{\exp(s(\mathbf{u}_i,\mathbf{u}_c))}{\sum_{n_j\in N_t}\exp(s(\mathbf{u}_i,\mathbf{u}_j))}
\end{equation}
where $\mathbf{u}_i$ represents the encoded representation vector of $n_i$, $s(\mathbf{u}_i,\mathbf{u}_j)$ is the similarity between node $n_i$ and $n_j$. $N_t$ represents the node set of type $t$ and $t= \phi(n_c)$.
The softmax function is normalized with respect to the node type of the context $n_c$, so each type of the neighborhood in the output layer is specified to one distinct multinomial distribution.

As we aim to embed each type $t\in \mathcal{T}$ of nodes into a distinct space,
we set $s(\mathbf{u}_i,\mathbf{u}_j) = \mathbf{u}_i^T \mathbf{u}_j$ if $n_i$ and $n_j$ have same type and in the same embedding space.
For each node pair $(\mathbf{u}_i,\mathbf{u}_j)$ with distinct node type $t_a$ and $t_b$ and connected by a relation with type $r$ (which can be an atomic relation or a composite relation), we formulate the similarity $s(\mathbf{u}_i,\mathbf{u}_j)$ by
%introduce two candidate relation-type-specific \PW{relation type specific? or node type specific? confusing} similarity metrics, which are described as:
\QZY{introduce two candidate similarity metrics specific for $r$, which are described as:}

\begin{itemize}
\item \textbf{Bilinear.} A multiplicative similarity metric. We introduce a bilinear metric $M_{r}\in \mathbb{R}^{d'\times d'}$, which is shared by each node pair $(\mathbf{u}_i,\mathbf{u}_j)$ with relation type $r$ and specific to different relation types to avoid the similarity calculation on respective semantics to dampen each other.

\begin{equation}
s(\mathbf{u}_i,\mathbf{u}_j)= {\mathbf{u}_i}^T M_{r} \mathbf{u}_j
\end{equation}

\item \textbf{Perceptron.} A additive similarity metric. We first input $\mathbf{u}_i$ and $\mathbf{u}_j$ to node-type-specific multilayer perceptrons with output metric dimension $d_m$, then add the outputs of them, next with a tanh activation layer we can obtain the hidden representation vector of node pair $(\mathbf{u}_i,\mathbf{u}_j)$. Then a relation-type-specific metric $m_r \in \mathbb{R}^{d_m}$ is multiplied to calculate the similarity.

\begin{equation}
s(\mathbf{u}_i,\mathbf{u}_j)= m_{r}^T \tanh(M_{t_a} \mathbf{u}_i + M_{t_b} \mathbf{u}_j)
\end{equation}

\end{itemize}

The metrics are trainable parameters and are learned to measure which $\mathbf{u}_j$ under the relation $r$ is closer to $\mathbf{u}_i$, whose mechanism is similar to the mechanism of attention models.
\QZY{Note that the parameters of metrics are shared for same relation type $r$, while different relation types make use of different metrics} and the corresponding distance on these relations of nodes is calculated in different metric spaces, %which helps to avoid the learned node representations interfered by different semantics.

Then, we adopt the popular negative sampling method proposed in \cite{mikolov2013distributed} to sample negative nodes to increase the optimization efficiency. Then, the probability $\log p(n_c|n_i,\theta)$ can be approximately defined as:

\begin{equation}
\label{eq6}
\log\sigma (s(\mathbf{u}_i,\mathbf{u}_c)) + \sum\nolimits_{n_j\in N_{neg}^i}\log\sigma(-s(\mathbf{u}_i,\mathbf{u}_j))
\end{equation}
where $\sigma(x)=\frac{1}{1+e^{-x}}$ is the sigmoid function and $N_{neg}^i \subseteq N_t$ is a negative node set for $n_i$ sampled from a pre-computed node frequency distribution $P_t(n_i)$, we set $P(n_i)\propto f_{n_i}^{3/4}$, where $f_{n_i}^{3/4}$ is the frequency of $n_i$ in $\mathcal{P}$.
Then, we conduct random walks on the $G$ to sample a set of paths $\mathcal{P}$.
Therefore, we can use skip-gram model on these paths and reformulate the objective in eq.\ref{objective} as follows:

\begin{equation}
\label{eq7}
\mathcal{L}=- \sum_{p\in \mathcal{P}}\sum_{n_i\in{p}}\sum_{n_c\in{C_i^k}}\log p(n_c|n_i,\theta)+\lambda{\|\theta\|}^2
\end{equation}
where $k$ is the windows size of context, $C_i^k$ is the context set containing the previous $k$ context nodes and next $k$ context nodes of $n_i$ in the path $p$.
Parameter $\lambda$ controls penalty of regularization for over-fitting. Finally, we use a mini-batch Adam optimizer to minimize $\mathcal{L}$ and optimize the parameters in the whole model.

\section{experiment}

In this section, We conduct several experiments to validate the performance of our proposed T-GNN model on three datasets.
The results demonstrate the superiority of our proposed method over several state-of-the-art methods. All codes of our method are publicly available\footnote{https://github.com/joe817/T-GNN}.

\subsection{Datasets}
To evaluate the proposed method, we use three kinds of real-world datasets:
%, which are widely used benchmarks for heterogeneous graph methods
academic graphs (DBLP), movie graphs (MOVIE), and review graph (YELP). The detailed descriptions of these datasets are as follows:

\begin{itemize}
\item \textbf{DBLP\footnote{https://www.aminer.cn/citation}:} DBLP is an academic bibliography with millions of publications. We extract a subset of DBLP with four areas: \emph{Data Mining (DM), Database (DB), Natural Language Processing (NLP)} and \emph{Computer Vision (CV)} from the year of 2013 to 2017. For each area, we choose three top venues\footnote{\emph{DM}: KDD, ICDM, WSDM. \emph{DB}: SIGMOD, VLDB, ICDE. \emph{NLP}: ACL, EMNLP, NAACL. \emph{CV}: CVPR, ICCV, ECCV.} and related papers(P), terms(T), authors(A) to construct an heterogeneous graph.

\item \textbf{IMDB\footnote{https://www.kaggle.com/tmdb/tmdb-movie-metadata}:} IMDB dataset contains knowledge about movies. We extract three genres of movie information from IMDB: \emph{Action, Comedy} and \emph{Drama}. Then we construct a movie heterogeneous graph which contains movies(M), actors(A), directors(D) and Tags(T).

\item \textbf{YELP\footnote{https://www.kaggle.com/yelp-dataset/yelp-dataset}:} YELP is a datasets containing the data of restaurants reviews. We extract a subset of YELP containing costumers(C), review(V), keywords(K) and restaurants(R) from the information of restaurants with types: \emph{American, Chinese, Japanese} and \emph{French} to form a restaurant review heterogeneous graph.
\end{itemize}

We select some types of nodes and use graph representation learning methods to learn their representation vectors,
Table \ref{SD} shows the statistics of these nodes and their corresponding hierarchical tree schemas used in our proposed model.
%Notice that there exists over-smoothing problems\cite{zhou2018graph} in traditional GNN models learning node representations, that is, with the increase of network layers or iterations in training, the hidden layer representation of nodes \QZY{in the same connected components of graph %\PW{what is ``in the same connected components''?}
%tend to converge to the same value (i.e. the same point in space), selecting different neighborhood ranges is a way to prevent over-smoothing\cite{Zhao2020PairNorm}.
%Therefore, we only choose acyclic hierarchical tree schemas to sample neighbors, so that in each extracted tree, nodes would not connect to those neighbors have the same type with it. For GNN model, nodes with same type are not in the same connected components, which alleviate the over-smoothing problem in their type-specific spaces.}

%As we embed node into type-specific spaces, in traning, these would not exist two node  with same type in same the connected component.
%each space there would not exist connected components, the node proximity are preserved via similar neighborhoods.}
%\PW{This paragraph is not clear. Especially the last sentence.}

\begin{table}[]
\setlength{\abovecaptionskip}{-0.05cm}
\caption{Statistics of three datasets}
\label{SD}
\begin{center}
\begin{tabular}{l|l|c}
\toprule
Datasets     & \multicolumn{1}{c|}{Node (\# Number)} & \begin{tabular}[c]{@{}l@{}}Hierarchical\\ Tree Schema\end{tabular} \\ \midrule
\multirow{3}{*}{DBLP} & Paper \#20,552    & $\overrightarrow{AP}$,\ $\overrightarrow{TP}$\        \\
                      & Author \#19,247   & $\overrightarrow{TPA}$                       \\
                      & Venue \#12       & $\overrightarrow{TPV}$,\ $\overrightarrow{APV}$                     \\ \midrule
\multirow{3}{*}{IMDB} & Movie \#3,630     & $\overrightarrow{AM}$,\ $\overrightarrow{TM}$                             \\
                      & Actor \#13,156    & $\overrightarrow{TMA}$                                                \\
                      & Director \#1,932  & $\overrightarrow{AMD}$,\ $\overrightarrow{TMD}$            \\ \midrule
\multirow{2}{*}{YELP} & Costumer \#2,536 & $\overrightarrow{KVC}$                                                \\
                      & Restaurant \#3,332 & $\overrightarrow{KVR}$                                       \\ \bottomrule
\end{tabular}
\end{center}
\end{table}

\subsection{Comparison Methods}
To validate the performance of our proposed method, we compare with some state-of-art baselines, including homogeneous heterogeneous graph representation learning method (DeepWalk), heterogeneous graph representation learning methods (Metapath2vec, RHINE), homogeneous graph neural networks (GraphSAGE, GAT) and heterogeneous graph neural networks (HAN, HetGNN). The brief descriptions of these baselines are as follows:

\begin{itemize}
\item\noindent\textbf{DeepWalk:}DeepWalk\cite{perozzi2014deepwalk} is a graph representation learning method based on random walks and skip-gram model to learn latent node representations.

\item\noindent\textbf{Metapath2vec:} Metapath2vec\cite{dong2017metapath2vec} uses meta-path based random walks to construct the heterogeneous neighborhood of nodes and then leverages a heterogeneous skip-gram model to perform node representations. We leverage the meta-paths APVPA, AMDMA and CRC in DBLP, IMDB, and Yelp respectively.

\item\noindent\textbf{RHINE:} RHINE\cite{lu2019relation} explore the different structural characteristics of relations in HINs and present two structure-related measures for two distinct categories of relations: IR and AR. We select the same relations in their paper on DBLP and YELP, for IMDB, we select AM, AMD as IR, and TM, TMD, TMA as AR.

\item\noindent\textbf{GraphSAGE:} GraphSAGE\cite{hamilton2017inductive} learn node representations through different aggregation function form local neighborhood of nodes. We use the GCN module as the aggregator of GraphSAGE.

\item\noindent\textbf{GAT:} GAT\cite{velivckovic2017graph} consider the attention mechanism and measure
impacts of different neighbors’ feature information by multi-head self-attention neural network.

\item\noindent\textbf{HAN:} HAN\cite{wang2019heterogeneous} leverages node level attention and semantic-level attention to respectively learn the importance of neighbors based on same meta-path and different meta-paths. We select meta-paths CRC, RCR for YELP, and meta-paths same as theirs for DBLP, IMDB.

\item\noindent\textbf{HetGNN:} HetGNN\cite{zhang2019heterogeneous} jointly considered heterogeneous neighbors sampling, node heterogeneous contents encoding, type based neighbors aggregation, and heterogeneous types combination.
\end{itemize}

%\subsection{Experiment Setting}
We use Par2vec\cite{le2014distributed} to pre-train the text content of nodes and set the dimension of initial feature vector as 128. For the proposed T-GNN, we set all the hidden layer dimensions and final representation dimension of nodes as 128. We use the Adam optimizer with a learning rate of $10^{-3}$ and the l2 penalty parameter $\lambda$ is $10^{-4}$. We set the windows size $k$ of skip-gram model as 2, the size of negative samples as 3 for all datasets, the similarity metric is chosen to Perceptron and the metric space dimension $d_m$ is 32. As GAT and HAN only mention semi-supervised objective function for node representations, for a fair comparison, we optimize their models same as our method to learn unsupervised node representations. For the baselines, we set the dimension of nodes in three datasets also as 128 and the other hyper-parameter setting are based on default values or the values specified in their own papers. All the experiments are repeated many times to make sure the results can reflect the performances of methods.

\begin{table}[htbp]
\setlength{\abovecaptionskip}{-0.05cm}
\caption{Results of Clustering}
\label{clustering}
\begin{center}
\begin{tabular}{lllllll}
\toprule
Dataset       & \multicolumn{2}{c}{DBLP} & \multicolumn{2}{c}{IMDB} & \multicolumn{2}{c}{YELP} \\ \midrule
Metric             & NMI        & ARI       & NMI        & ARI        & NMI        & ARI       \\ \midrule
DeepWalk     &0.735        &0.616       &0.023        &0.015        &0.260        &0.279    \\
Metapath2vec &0.864        &0.899       &0.096        &0.091        &0.261        &0.282     \\
RHINE        &0.866        &0.902       &0.055        &0.036        &0.342        &0.351      \\ \hline
GraphSAGE    &0.865        &0.912       &0.128        &0.135        &0.396        &0.433     \\
GAT          &0.855        &0.897       &0.114        &0.113        &0.413        &0.457        \\
HAN          &0.900        &0.933       &0.136        &0.144        &0.413        &0.458            \\
HetGNN       &0.891        &0.939       &0.131        &0.139        &0.403        &0.440      \\ \hline
T-GNN      &\textbf{0.916}&\textbf{0.955}&\textbf{0.145}&\textbf{0.152}&\textbf{0.420}&\textbf{0.484 }\\ \bottomrule
\end{tabular}
\end{center}
\end{table}

\begin{table}[htbp]
\caption{Results of Multi-class Classification}
\label{classification}
\begin{tabular}{lllllll}
\toprule
Dataset       & \multicolumn{2}{c}{DBLP} & \multicolumn{2}{c}{IMDB} & \multicolumn{2}{c}{YELP} \\ \midrule
Metric(F1)   &Micro        & Macro      &Micro        & Macro       &Micro        & Macro       \\ \midrule
DeepWalk     &0.905        &0.896       &0.478        &0.473        &0.703        &0.660       \\
Metapath2vec &0.922        &0.918       &0.505        &0.509        &0.705        &0.660       \\
RHINE        &0.927        &0.920       &0.449        &0.448        &0.726        &0.676       \\ \hline
GraphSAGE    &0.962        &0.943       &0.583        &0.584        &0.739        &0.748       \\
GAT          &0.967        &0.958       &0.543        &0.542        &0.744        &0.758            \\
HAN          &0.979        &0.971       &0.596        &0.598        &0.746        &0.759       \\
HetGNN       &0.983        &0.979       &0.594        &0.593        &0.730        &0.753         \\ \hline
T-GNN      &\textbf{0.997}&\textbf{0.996}&\textbf{0.608}&\textbf{0.609}&\textbf{0.760}&\textbf{0.772}       \\ \bottomrule
\end{tabular}
\end{table}

\subsection{Clustering and Classification}
First, we conduct clustering and classification tasks to evaluate the performance of the methods.
%\PW{What do you mean by ``we evaluate XXX labeled by xxx''. Do you mean to express ``define the types of nodes based on some criterion''?}
\QZY{For DBLP, we label papers by areas of their venues and label authors by their representative areas (the area with the majority of their papers). For IMDB, we label the movies by genres and label actors by the genre with the majority of their acted movies. For YELP, we label the restaurant by type. Then we evaluate these nodes' representations by clustering and multi-class classification.}
%and customers labeled by the majority one of their reviewed restaurants types.
We learn the node representations on the heterogeneous graphs by each model. For clustering task, we feed node representations into the K-Means algorithm, the number of clusters is set to the true number of classes for each dataset, we evaluate the clustering performance in terms of normalized mutual information (NMI) and adjusted rand index (ARI). For multi-class classification task, the learned node representations are used as the input to a logistic regression classifier, the proportion of training and valid data is set to 50\%, 10\%, the remaining nodes are used for test. We use both Micro-F1 and Macro-F1 as classification evaluation metrics.

\textbf{Results.}
Table \ref{clustering} shows the performance evaluation of clustering on three datasets and Table \ref{classification} shows the classification results. The proposed T-GNN model consistently outperform all baselines in terms of two tasks, showing that it can learn more effective node representations than other methods. The relative improvements (\%) of T-GNN over the best baselines range from 1.7\%-10.7\% for clustering task and 1.4\%-2.2\% for classification task.
%\PW{you use ``graph'' and ``network'' interchangeably. Please unify.}
\QZY{ We can also observe 1) GNN based graph representation learning models achieve better performance than the traditional graph representation learning methods on these two tasks, that is because besides the graph structure information, GNNs can also capture node attributes information to node representations.
2) Heterogeneous models perform better than homogeneous models because they can differentiate different types of relations and capture more semantic information to node representations.
3) T-GNN's performance is better than two heterogeneous graph neural network models HAN and HetGNN, showing that it can better capture the information of heterogeneous neighborhood.}

\begin{figure}[htbp]
\setlength{\belowcaptionskip}{-0.3cm}
\centering

\subfigure[Inductive Clustering (10\%)]{
\begin{minipage}[t]{0.48\linewidth}
\centering
\includegraphics[width=1\linewidth]{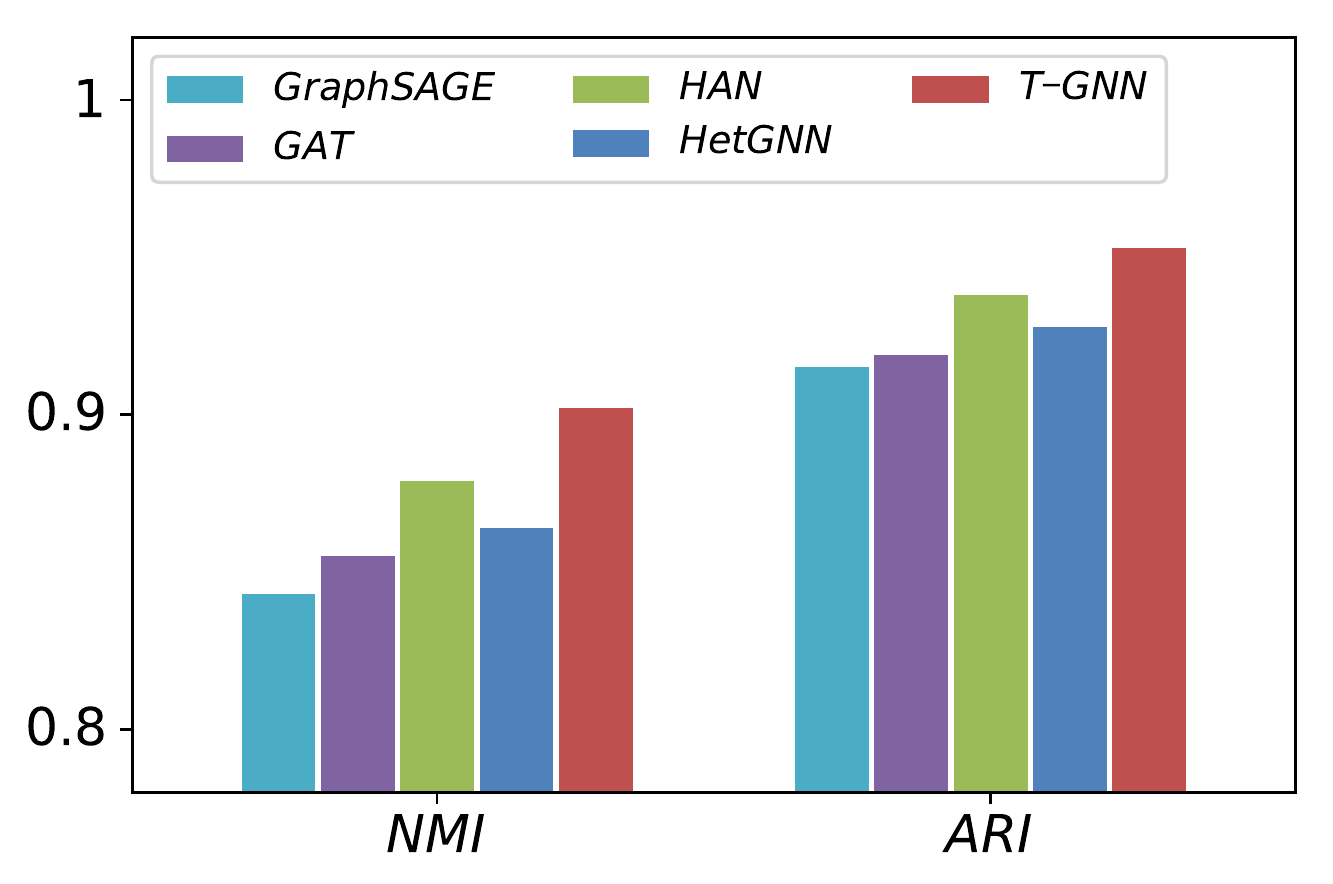}
%\caption{}
\end{minipage}%
}%
\subfigure[Inductive Clustering (20\%)]{
\begin{minipage}[t]{0.48\linewidth}
\centering
\includegraphics[width=1\linewidth]{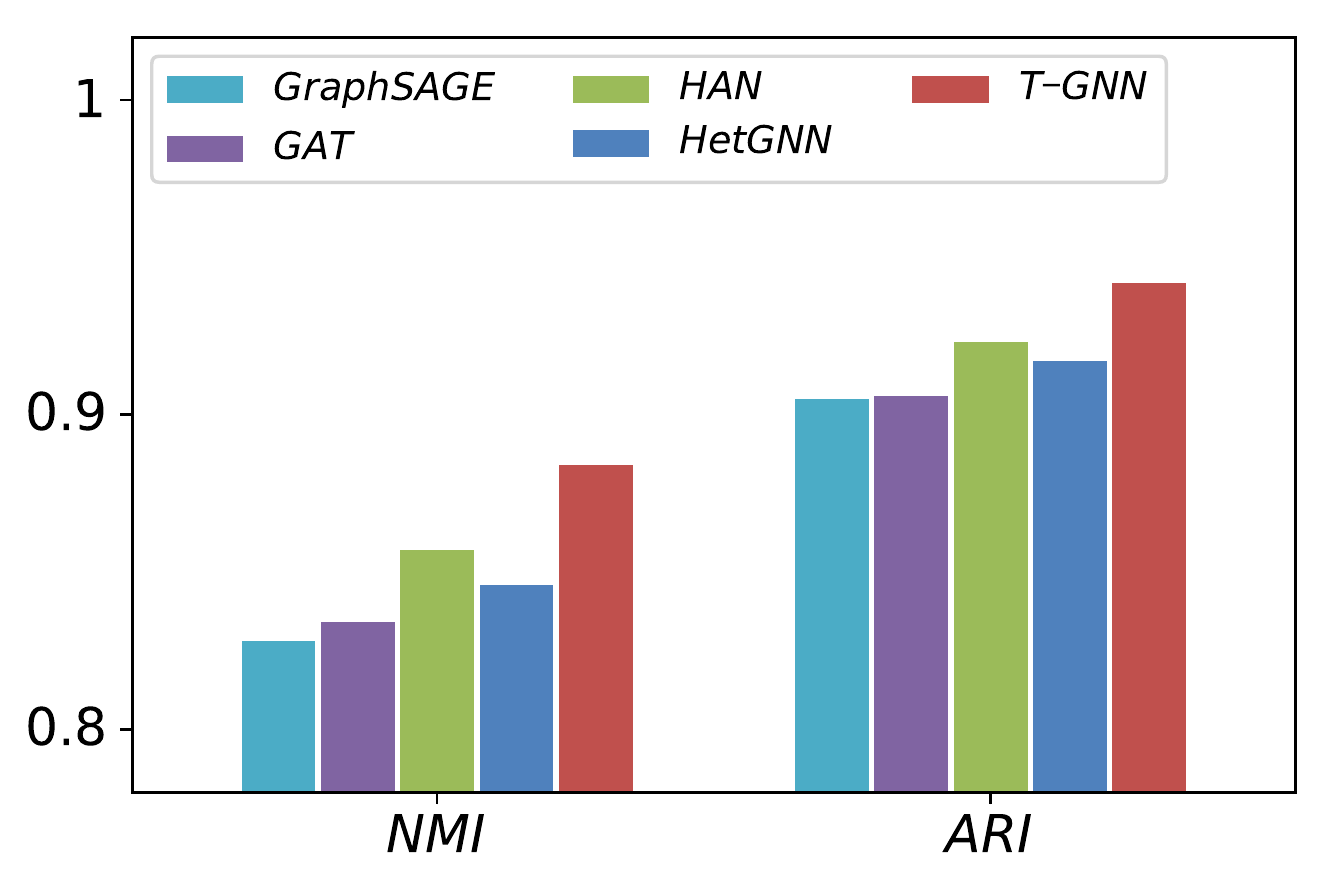}
%\caption{}
\end{minipage}%
}%
\quad                 %这个回车键很重要 \quad也可以
\subfigure[Inductive Classification (10\%)]{
\begin{minipage}[t]{0.48\linewidth}
\centering
\includegraphics[width=1\linewidth]{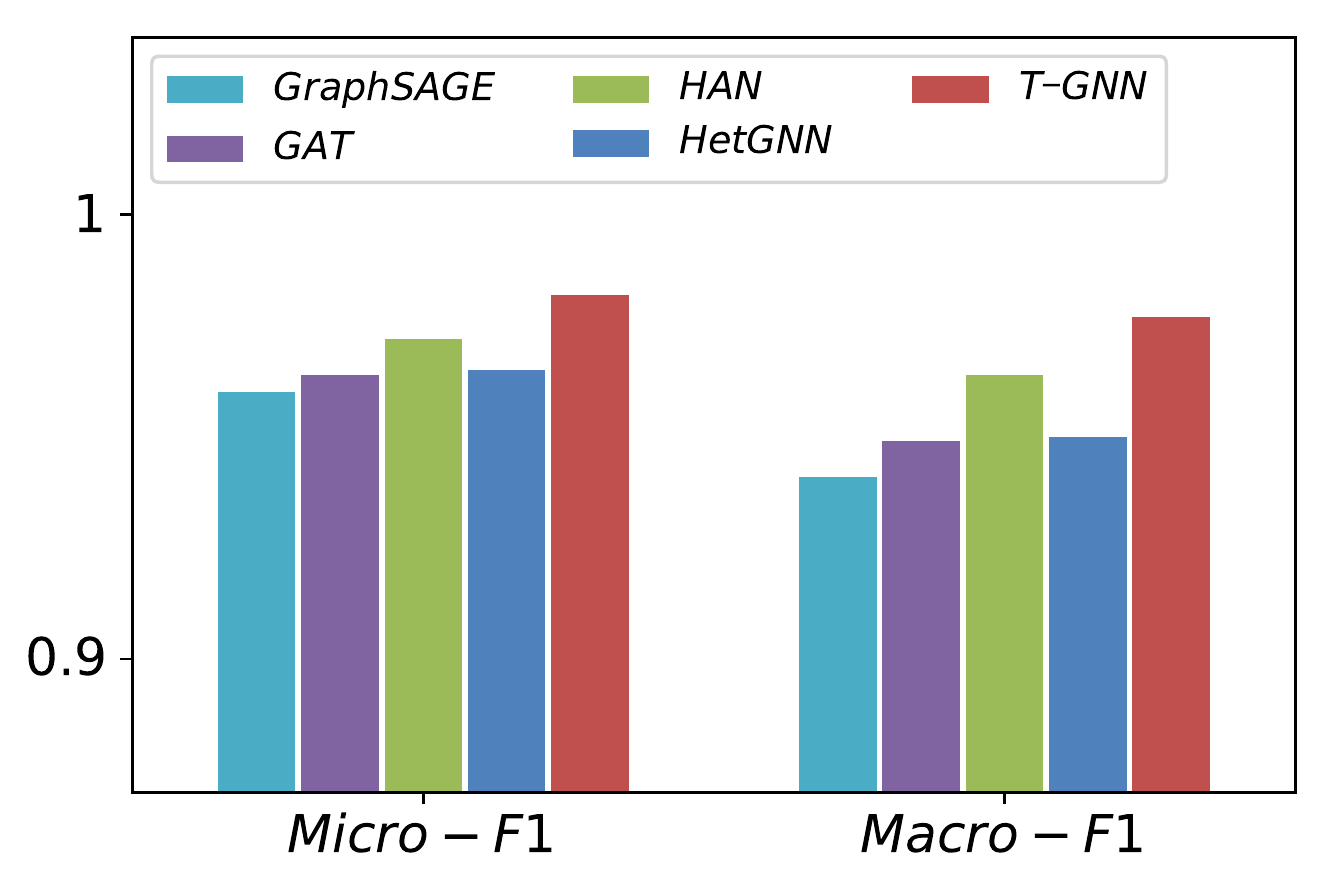}
%\caption{fig2}
\end{minipage}
}%
\subfigure[Inductive Classification (20\%)]{
\begin{minipage}[t]{0.48\linewidth}
\centering
\includegraphics[width=1\linewidth]{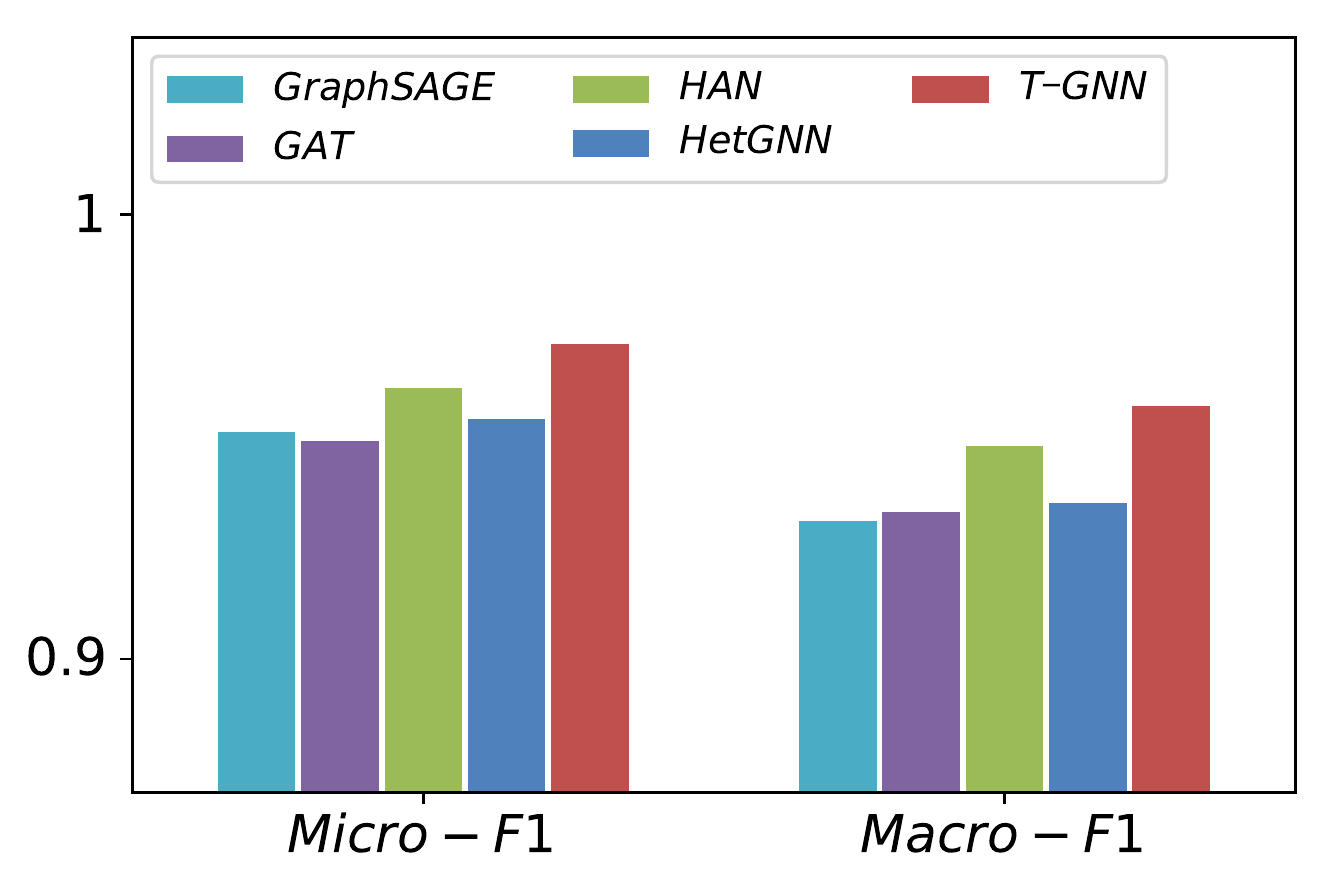}
%\caption{fig2}
\end{minipage}
}%

\centering
\caption{Results of inductive clustering and inductive multi-class classification, where the percentages in captions represent the invisible ratios of nodes.}
\label{ICC}
\end{figure}

\subsection{Inductive Clustering and Classification}
% As the parameters in GNN based models are independent of the number of nodes in graphs and can be utilized on inductive tasks.
In this task, we evaluate the inductive capability of our proposed T-GNN on DBLP dataset and compare with the graph neural network based baselines GraphSAGE, GAT, HAN and HetGNN. Specifically, for author nodes and paper nodes which to be evaluated, we randomly set a part of them as the invisible nodes for graph representation learning methods, and use the rest visible nodes and the links between them on the graph to train the models.
Then we use the learned models on the whole graph to infer the representations of all invisible nodes.
Finally, we use the inferred representations as the input of clustering tasks.
For the classification tasks, the visible nodes are used to train the classifier, and the invisible nodes are used to evaluate.
The labels and evaluation metrics are the same as the previous clustering and classification tasks.
We report the results of the invisible nodes ratio in as 10\% and 20\% respectively.

\textbf{Results.}
Figure \ref{ICC} shows the performance of GNN models on the inductive learning task.
% In these two tasks, all methods shows promising performance in inferring representations of invisible nodes.
Our proposed T-GNN model achieves the best performance among all the homogeneous and heterogeneous models.
Specifically, for the comparison of heterogeneous models, our proposed T-GNN outperforms HAN and HetGNN, because the representations of invisible nodes are mainly inferred by their neighborhood information, but HAN uses meta-path to sample neighbors, ignoring the intermediate nodes; HetGNN uses random walks to sample neighbors, them both unable to process tree structured information and path correlation on multi-hop neighborhood.
%ignoring the sequentiality of neighbor information; and they both ignore the structural property of neighborhoods.

\subsection{Link Prediction}
In this section, we aim to evaluate the performance of our graph representation learning methods on the link prediction task, which is a practical problem in many applications such as user/item recommendation. We formulate this task as a binary classification problem to predict whether a relation link between two nodes exists. We consider two types of links: links of two author collaborating one paper in DBLP dataset, and links of a user giving a review to an restaurant in YELP dataset. Specifically, for DBLP, we sample the collaborating links before 2013 for training, in 2013 for validation and after 2013 for test.
For YELP, we randomly sample 60\% reviewing links for training, 10\% for validation, and the rest for testing.
In training, we remove the links in the validation and testing sets and use graph representation learning methods on graphs to learn the node representations.
We use the element-wise multiplication of two candidate nodes' representations as the link representation, then input link representations into a binary logistic classifier. Also negative links (not connected in graphs) with three times the number of true links are randomly sampled to the training, validation and testing sets. We use AUC and F1 as evaluate metrics.

\textbf{Results.}  Table \ref{LP} shows the prediction results on two datasets. We can find that our proposed T-GNN model achieves the best performance or is comparable to the best methods on the link prediction task.
%\PW{split the last sentence into several short sentences. You do not need to write long and complex sentences. Clear writing is the first priority.}
\QZY{We believe the reason is that our proposed T-GNN model has outstanding ability in preserve heterogeneity into node representations, which helps to overcome the mutual interference of different relations on the node distributions, and improve the quality of node representations.}

\begin{table}[htbp]
\setlength{\abovecaptionskip}{-0.05cm}
\caption{Results of Link Prediction}
\label{LP}
\begin{center}
\begin{tabular}{lllll}
\toprule
Dataset       & \multicolumn{2}{c}{\begin{tabular}[c]{@{}c@{}}DBLP\\ (Collaborating)\end{tabular}} & \multicolumn{2}{c}{\begin{tabular}[c]{@{}c@{}}YELP\\ (Reviewing)\end{tabular}} \\ \midrule
Metric             & AUC        & F1      & AUC        & F1       \\ \midrule
DeepWalk     & 0.691       & 0.539      &0.519        &0.392       \\
Metapath2vec & 0.723       & 0.622      &0.518        &0.390       \\
RHINE        & 0.729       & 0.630      &0.552        &0.410       \\ \hline
GraphSAGE    & 0.713       & 0.584      &\textbf{0.665}&0.536       \\
GAT          & 0.719       & 0.603      &0.652        &0.550       \\
HAN          & 0.743       & 0.635      &0.655        &0.551       \\
HetGNN       & 0.760       & 0.660      &0.646        &0.536           \\ \hline
T-GNN      &\textbf{0.824} &\textbf{0.764} &0.663 &\textbf{0.571}      \\ \bottomrule
\end{tabular}
\end{center}
\end{table}

\subsection{Comparison of Variant Models}
\begin{figure}[htbp]
\setlength{\belowcaptionskip}{-0.3cm}
\centering
\subfigure[Clustering]{
\label{fig:knf} %% label for first subfigure
\begin{minipage}[t]{0.49    \linewidth}
\centering
\includegraphics[width=1\linewidth]{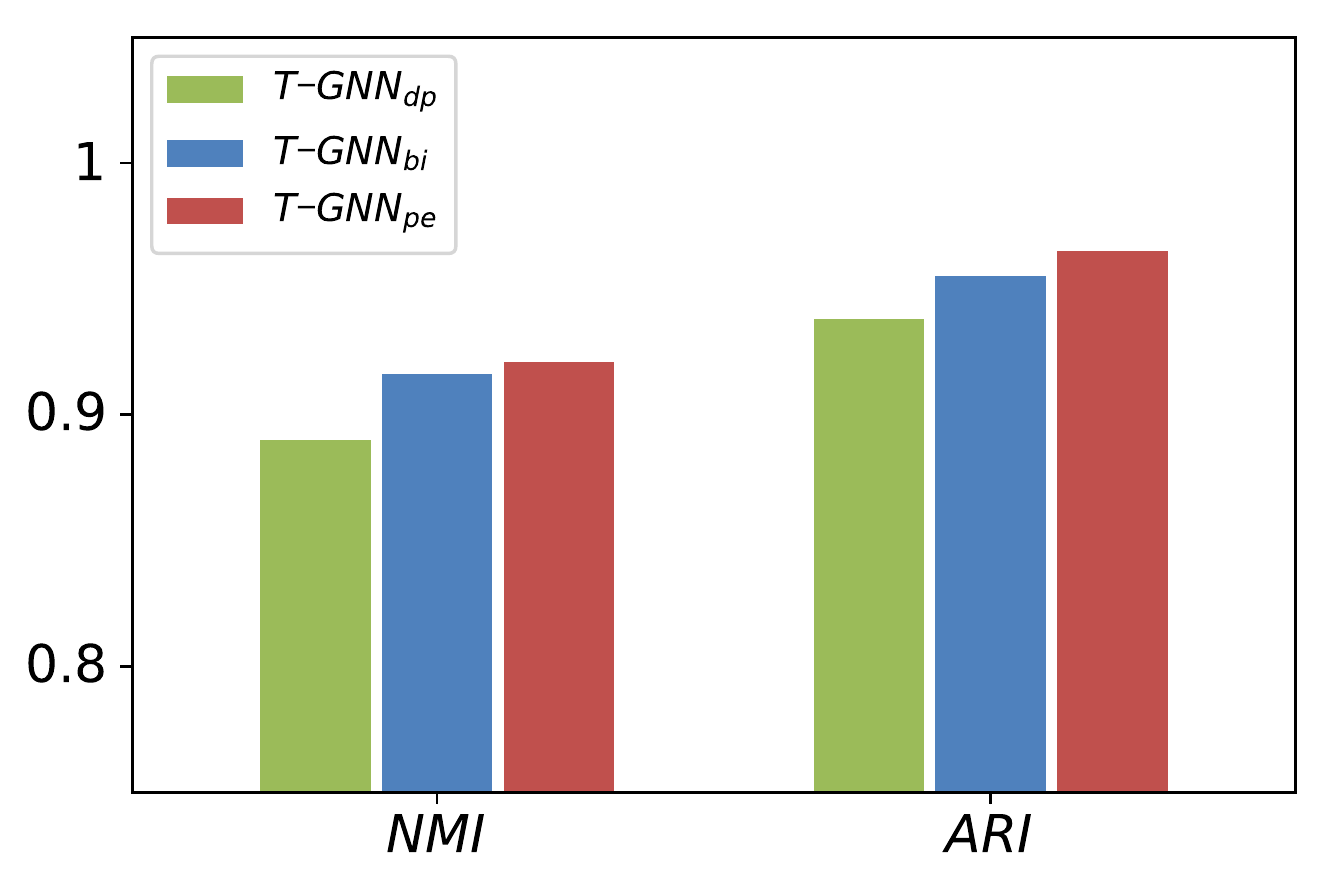}
\end{minipage}%
}%
\subfigure[Classification]{
\label{fig:kunf} %% label for second subfigure
\begin{minipage}[t]{0.49\linewidth}
\centering
\includegraphics[width=1\linewidth]{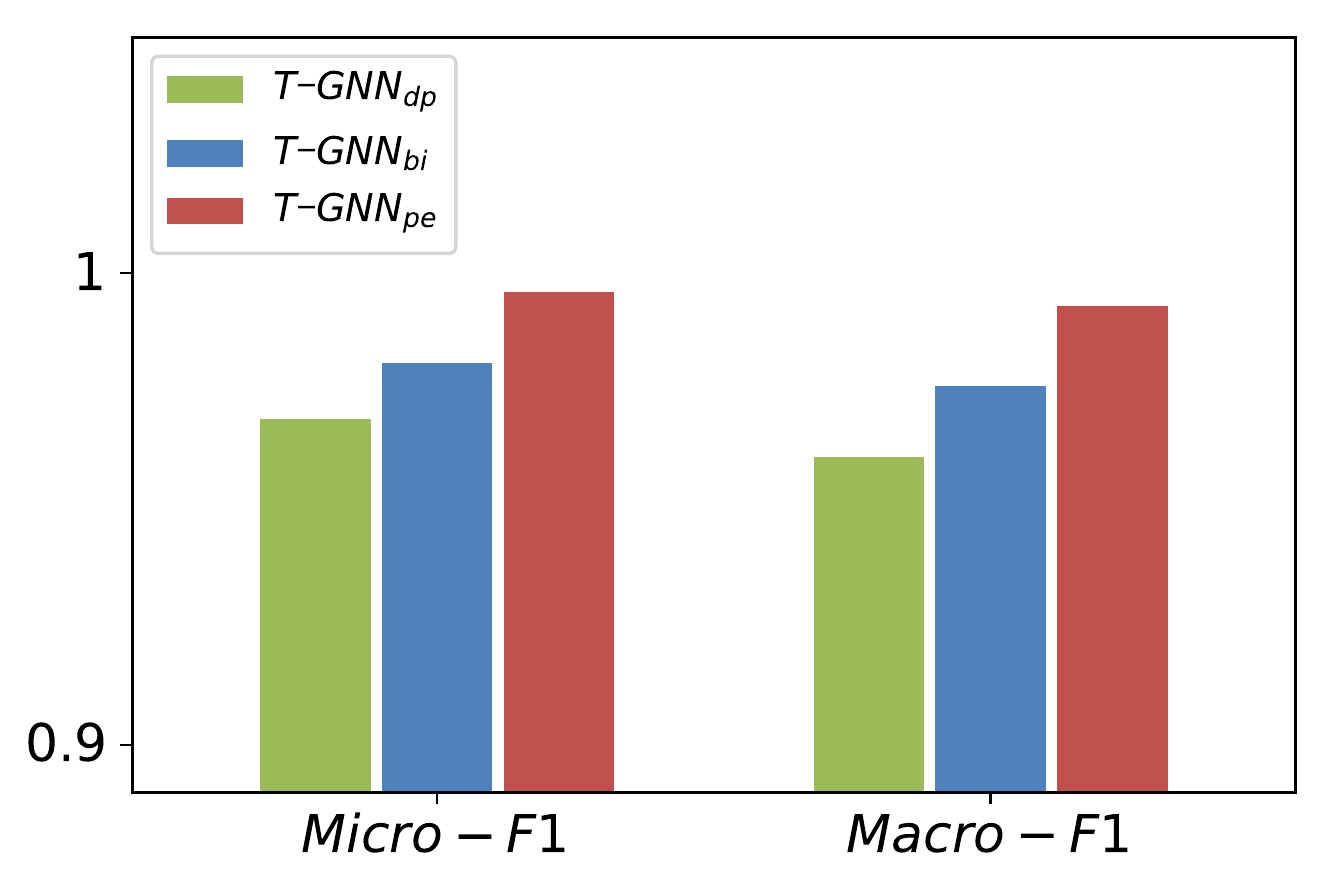}
\end{minipage}%
}%

\caption{Performance Evaluation of Variant Models.}
\label{fig:VM} %% label for entire figure
\end{figure}

In order to further verify the effectiveness of the metrics on learning node representations, we design three variant heterogeneous graph representation learning models based on T-GNN and different output models: T-GNN$_{dp}$, T-GNN$_{pe}$, T-GNN$_{bi}$. T-GNN$_{dp}$ do not use metrics and leverage dot product as the similarity $s(n_i,n_j)$ of any two nodes $n_i$ and $n_j$, T-GNN$_{pe}$ use the Perceptron as the similarity metric of nodes with different types and T-GNN$_{bi}$ use Bilinear, both of them are introduced in Section \ref{OM}. We evaluate these three models on the node clustering task and classification task.
The experiment setting are same as mentioned above. The results are shown in Figure \ref{fig:VM}.
It is evident that using two kinds of metrics achieves better performance.
Besides, the performance of T-GNN$_{pe}$ is better than T-GNN$_{bi}$.
We believe the reason is that there are more parameters in Perceptron metric than Bilinear metric, leading to Perceptron metric is better in preserve the complicated proximities between multiple nodes.

\begin{figure*}[htbp]
\setlength{\belowcaptionskip}{-0.3cm}
\centering
\subfigure[Metapath2vec (Author)]{
\label{fig:a} %% label for first subfigure
\begin{minipage}[t]{0.33\linewidth}
\centering
\includegraphics[width=1\linewidth]{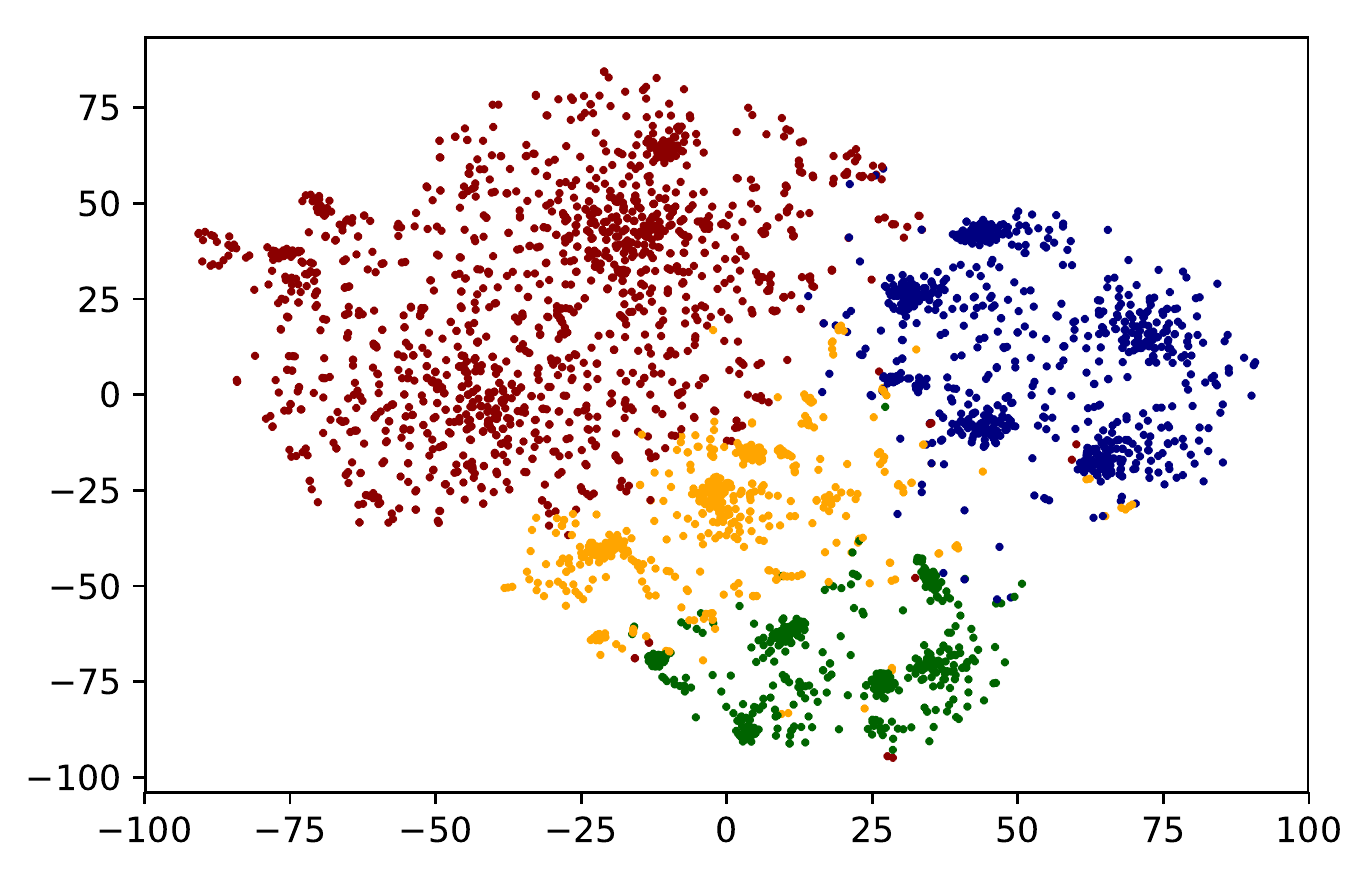}
\end{minipage}%
}%
\subfigure[GraphSAGE (Author)]{
\label{fig:b} %% label for second subfigure
\begin{minipage}[t]{0.33\linewidth}
\centering
\includegraphics[width=1\linewidth]{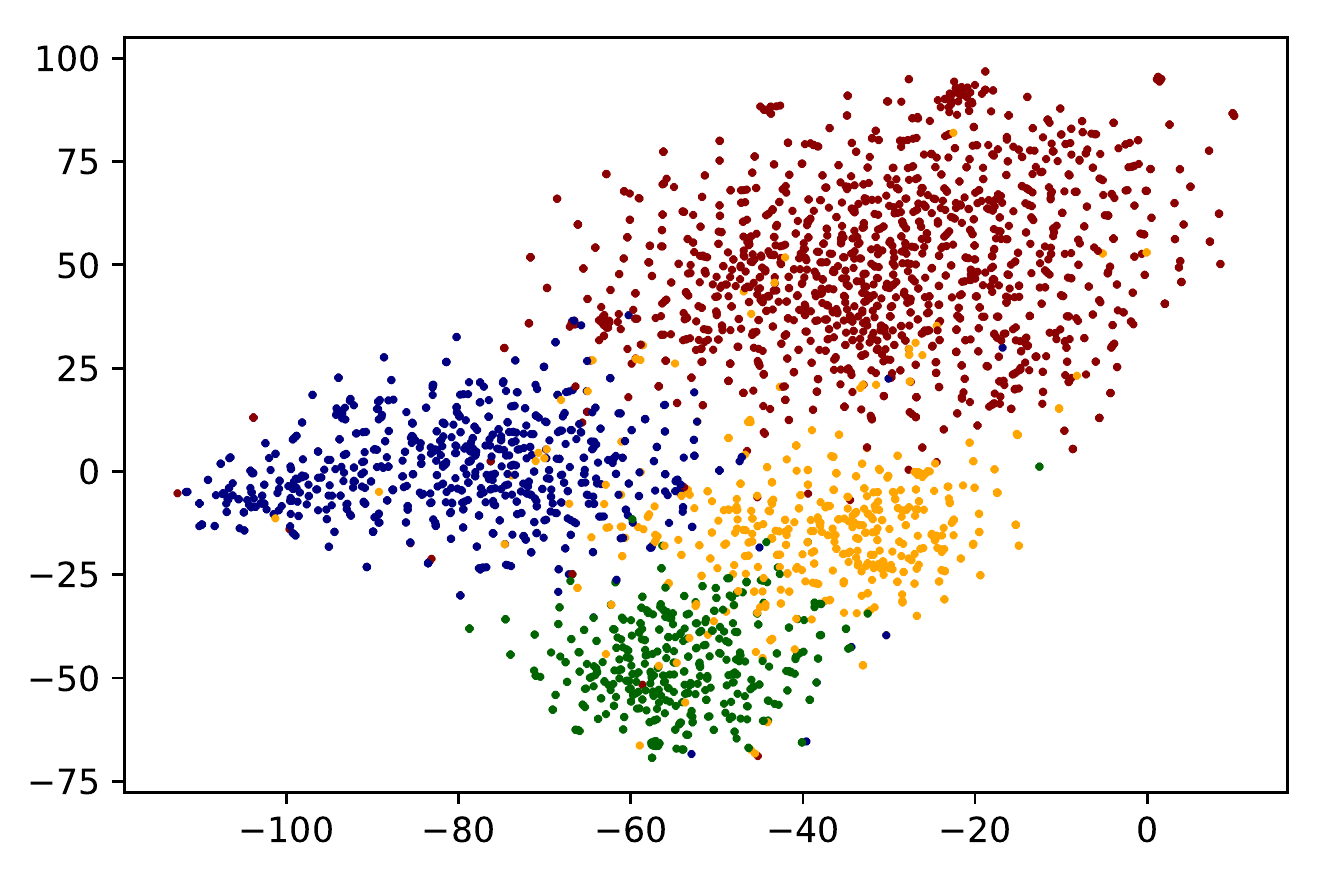}
\end{minipage}%
}%
\subfigure[T-GNN (Author)]{
\label{fig:c} %% label for second subfigure
\begin{minipage}[t]{0.33\linewidth}
\centering
\includegraphics[width=1\linewidth]{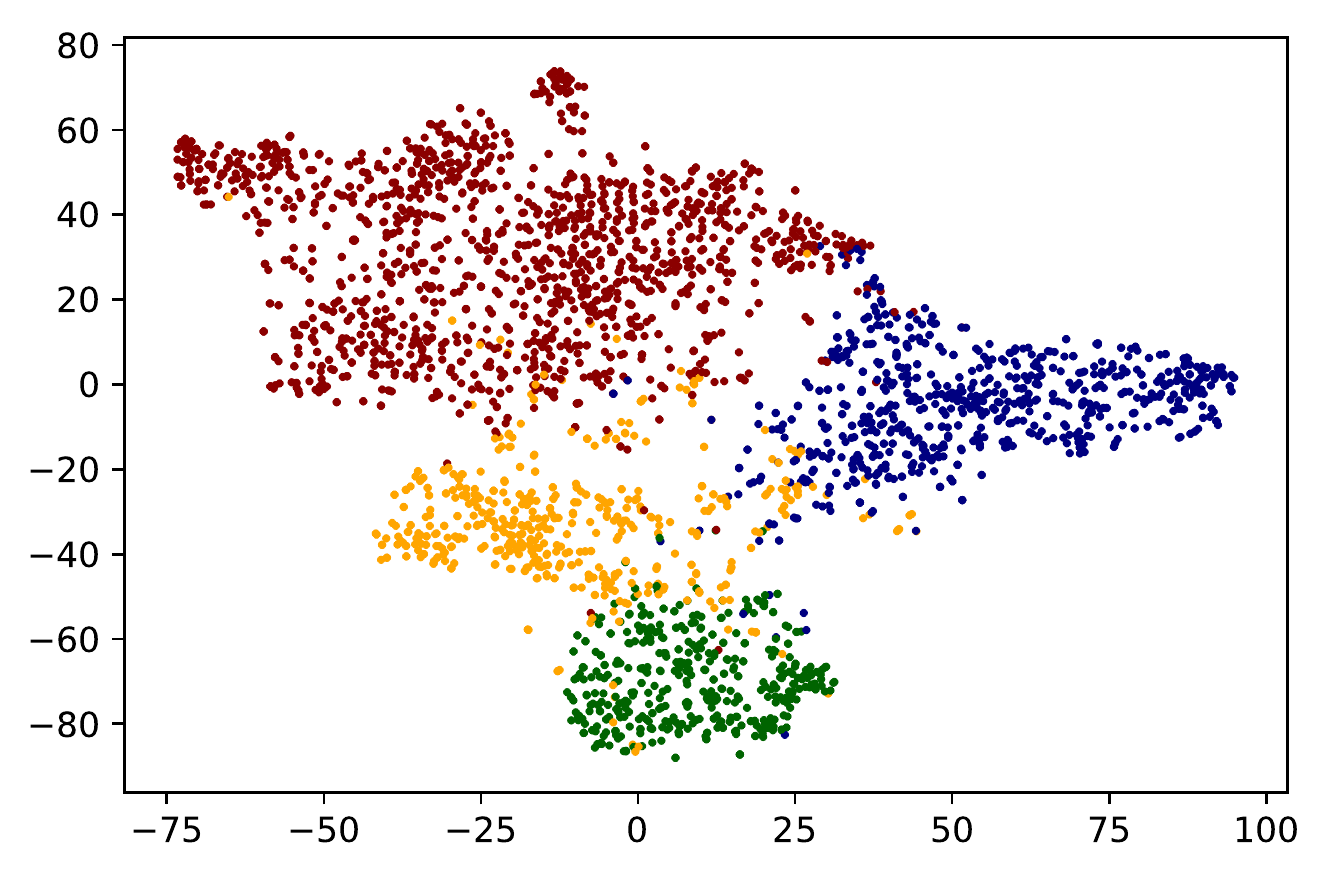}
\end{minipage}%
}%

\subfigure[Metapath2vec (Paper)]{
\label{fig:d} %% label for first subfigure
\begin{minipage}[t]{0.33\linewidth}
\centering
\includegraphics[width=1\linewidth]{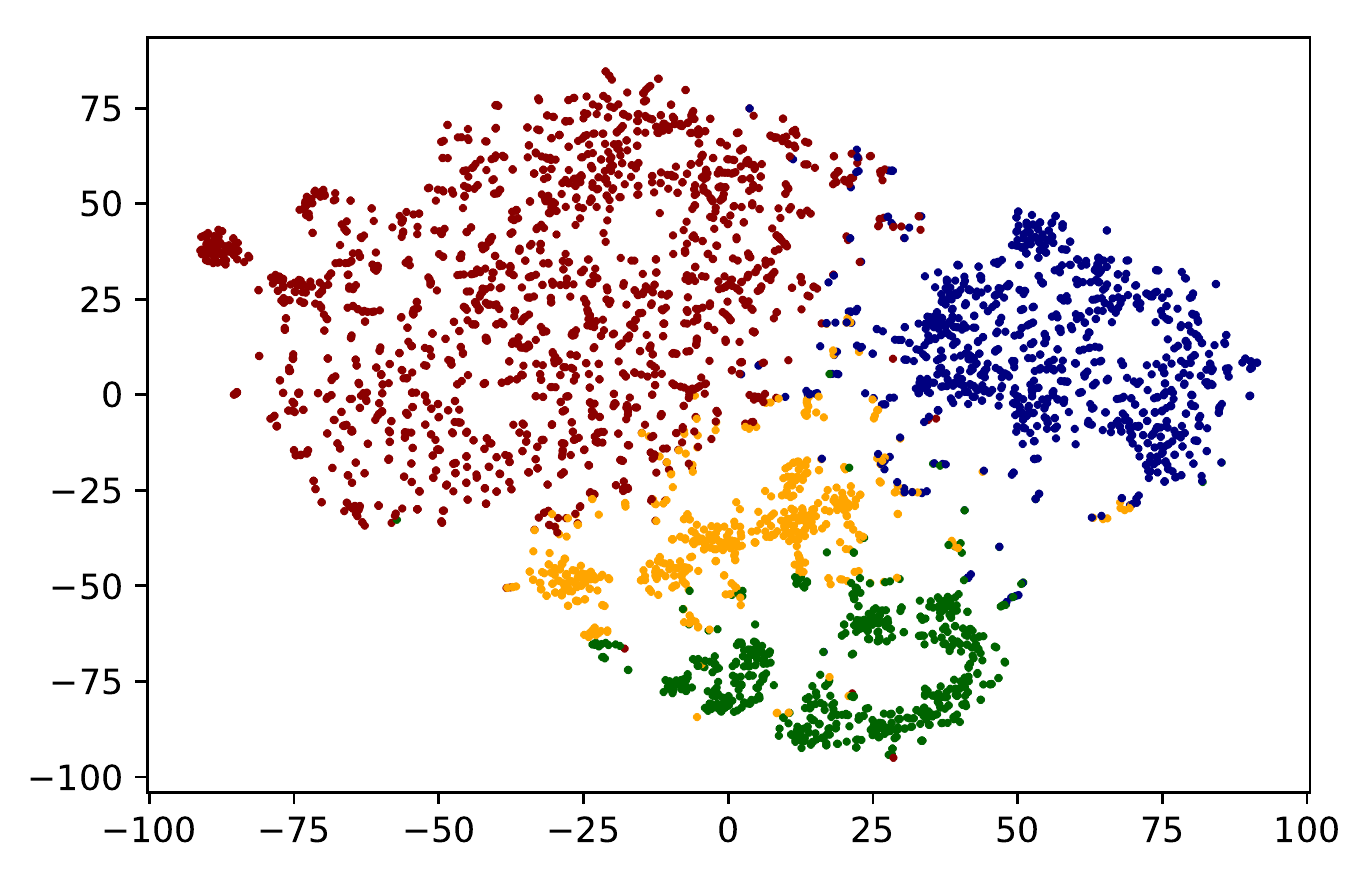}
\end{minipage}%
}%
\subfigure[GraphSAGE (Paper)]{
\label{fig:e} %% label for second subfigure
\begin{minipage}[t]{0.33\linewidth}
\centering
\includegraphics[width=1\linewidth]{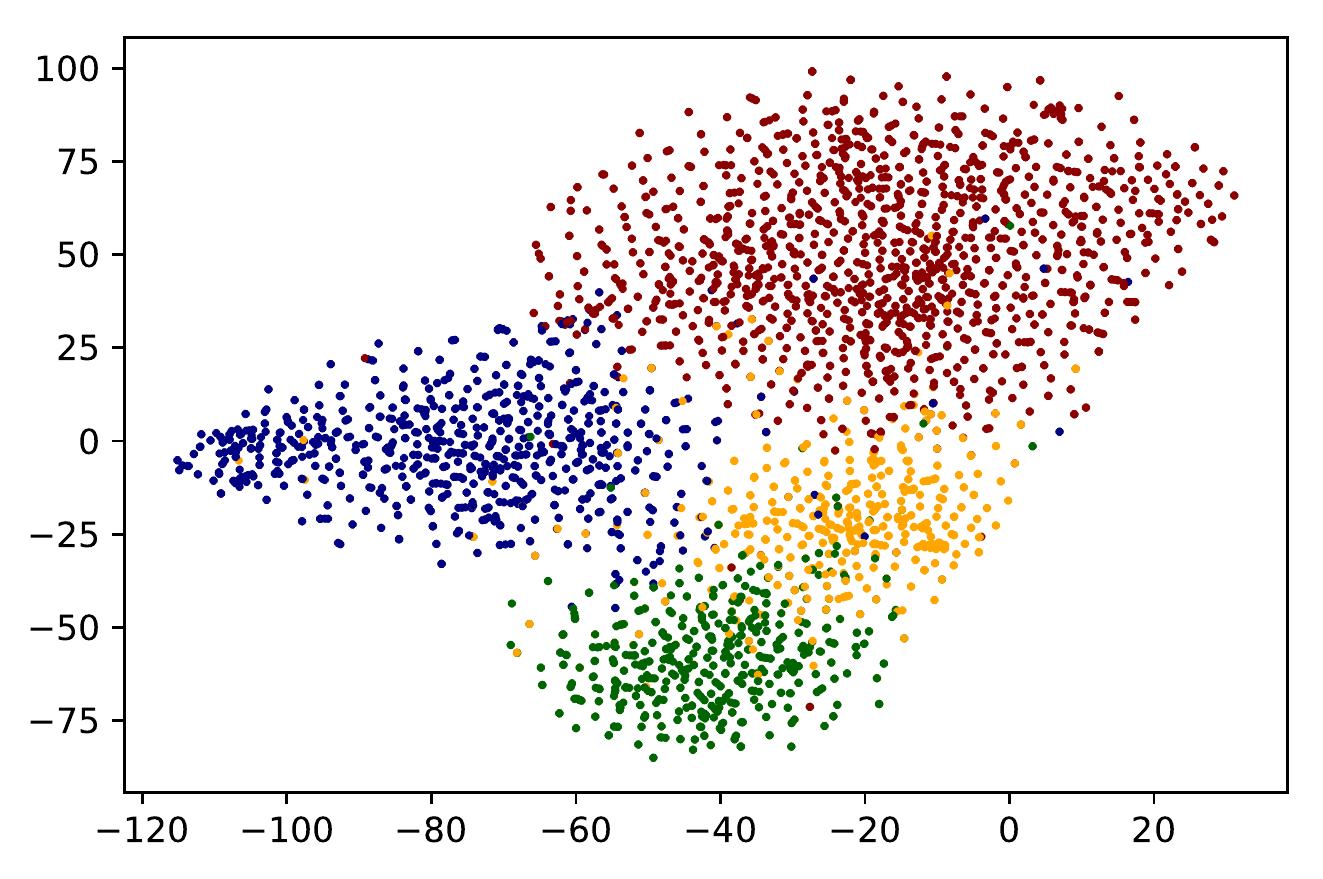}
\end{minipage}%
}%
\subfigure[T-GNN (Paper)]{
\label{fig:f} %% label for second subfigure
\begin{minipage}[t]{0.33\linewidth}
\centering
\includegraphics[width=1\linewidth]{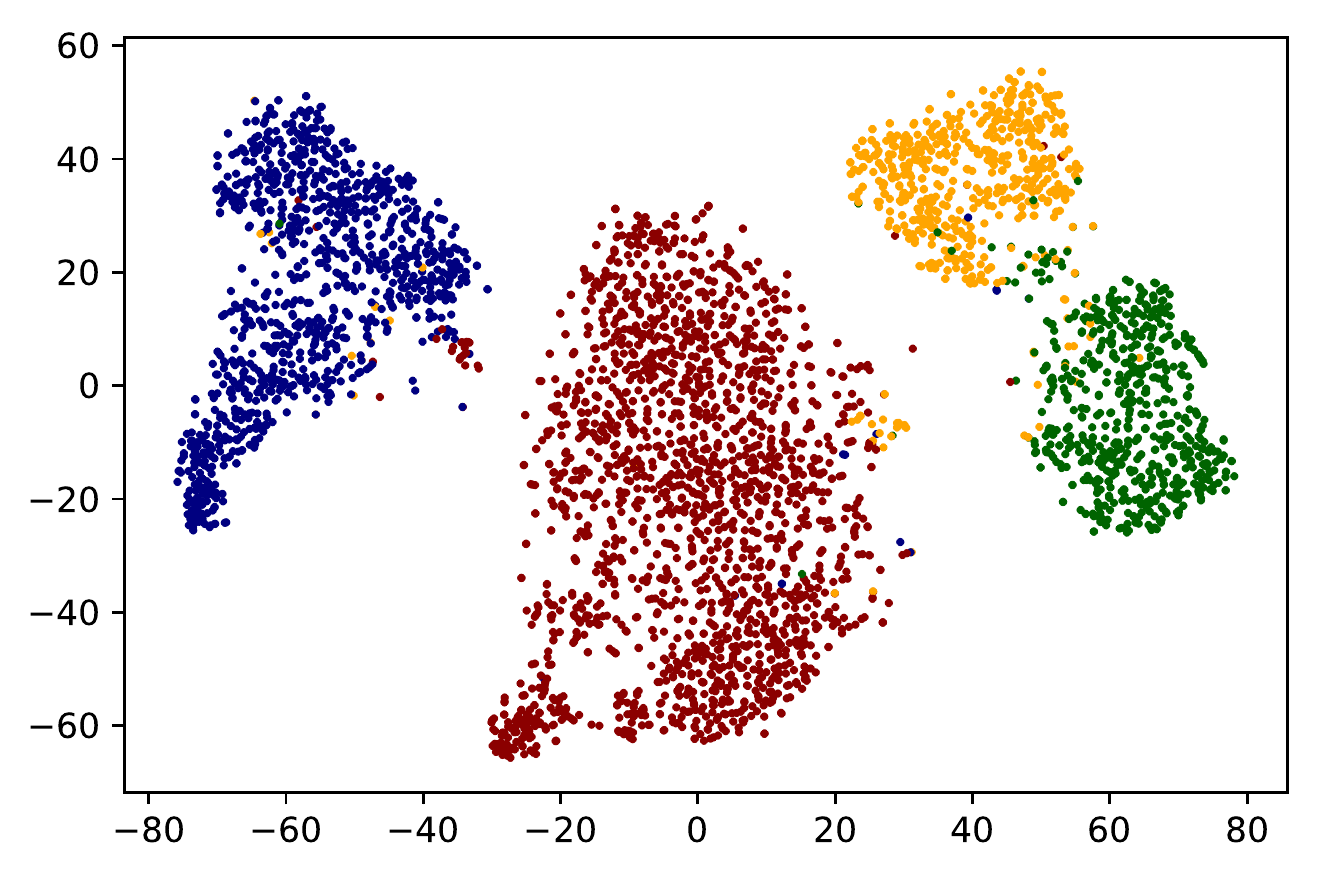}
\end{minipage}%
}%
\centering
\caption{t-SNE Visualization of representation distribution of authors and papers learned by three different graph representation learning methods: Metapath2vec, GraphSAGE and T-GNN. (a), (b), (c) denotes the author representations learned by these three methods and (d), (e), (f) denotes the paper representations.
The four different research areas which authors and papers belong to are colored differently.}
\label{fig:tsne} %% label for entire figure
\end{figure*}

\subsection{Visualization}
We have discussed that our model can embed each type of nodes into type-specific space with distinct distribution. In this section, we compare our model with metapath2vec and GraphSAGE by visualizing node representations with different types. Specifically, we first learn the node representations in the year of 2015 of four venue: CVPR, ACL, KDD, SIGMOD in DBLP data, representing four research areas. Then we use t-SNE to respectively project author representations and paper representations into 2-dimensional spaces.

The results are shown in Figure~\ref{fig:tsne}.
We can observe that while metapath2vec cluster papers and authors into small groups, some nodes belonging to the same research area are far away from each other, and some belonging to different areas are mixed with each other, GraphSAGE can roughly partition nodes with different research areas, but the boundaries between different clusters are not sharp, and many nodes overlap at the same points. We can observe that in the visualization of T-GNN,\QZY{ 1) the representations of paper nodes in the same area cluster closely and can be well distinguished from each other. Author nodes are also embedded well comparing with other two methods, but some of them are embedded on the boundary between different clusters, which is because some authors may have multiple research areas. 2) Authors and papers in their own space have independent distributions, showing high intra-class similarities and distinct boundaries between different research areas.}

%In addition, with the help of metrics, paper nodes and author nodes are embedded into different spaces, making they would not interfere each other in same space, so the learned representations with same type have independent distribution, showing high intra-class similarities and distinct boundaries between different research areas.

\section{Related work}
\textbf{Graph Representation Learning}
Graph representation learning, also known as network embedding, has become one of the most popular research interests in data mining recently. Network embedding method aim to embed network into lower dimensional space which preserve the network property and node proximity, the learned node embeddings can be utilized in many graph mining tasks. In homogeneous networks embedding methods, DeepWalk\cite{perozzi2014deepwalk} and Node2Vec\cite{grover2016node2vec} use random walk strategy on network and skip-gram\cite{mikolov2013efficient,rong2014word2vec} model to learn the representation of each node in network. LINE\cite{tang2015line} aims to learn the node embedding that preserve both first-order and second-order proximities. Some use deep neural network for homogeneous network embedding, such as DNGR\cite{cao2016deep} and SDNE\cite{wang2016structural}. Different with homogeneous graph, heterogeneous information network contains multiple types of nodes and relations, many real world graph can be modeled as HINs. Metapath2Vec\cite{dong2017metapath2vec} designs a meta-path based random walk and utilizes
skip-gram to perform heterogeneous graph embedding, HINE\cite{huang2017heterogeneous} define an objective function modeling the meta path based proximities in the embedded low-dimensional space. HIN2Vec\cite{fu2017hin2vec} learns the embeddings of HINs by conducting multiple prediction training tasks jointly, RHINE\cite{lu2019relation} explore the relation characteristics in HINs and partition them into two categories, then use different models to optimize the node representations. Besides, Some methods introduce metrics learning into heterogeneous network embedding, HEER\cite{shi2018easing} embeds HINs via edge representations that are further coupled with properly-learned heterogeneous metrics, PME\cite{chen2018pme} models node and link heterogenities in elaborately designed relation-specific spaces.

\textbf{Graph Neural Network}
Graph neural networks extend the deep neural network to encode the node features and local neighborhood into low-dimensional representation vectors\cite{zhou2018graph}. Kipf et al. propose GCN\cite{kipf2016semi}, which designs a graph convolutional network via a localized first-order approximation of spectral graph convolutions. GraphSAGE\cite{hamilton2017inductive} propose an general aggregating strategy containing multiple aggregators, such as Mean aggregator, LSTM aggregator and Pooling aggregator. For heterogeneous graph, R-GCN\cite{schlichtkrull2018modeling} introduces a relational graph convolutional network to link prediction task and entity classification task. Zhang et al. propose a heterogeneous graph neural network model HetGNN\cite{zhang2019heterogeneous} which considers both types and heterogeneous attributes of nodes. RSHN \cite{zhu2019relation} utilize graph structure and implicit relation structural information to simultaneously learn node and edge type embedding. Some work introduce attention mechanisms into network architectures, GAT\cite{velivckovic2017graph} employs self-attention mechanism to measure impacts of different neighbors and combine their impacts to obtain node representations. HAN\cite{wang2019heterogeneous} employs node-level and semantic-level attentions on heterogeneous network  to learn the importance of neighbors based on meta-paths. Also some work introduce gated recurrent units, GGNN\cite{li2015gated} aim learning representations of the internal state during the process of producing a sequence of outputs. Inspired by this idea, \cite{beck2018graph} proposed a novel encoder-decoder architecture for graph-to-sequence learning. \cite{gilmer2017neural} propose general conceptual message passing neural networks framework that subsumes most GNNs and evaluate it on molecular property prediction task. \cite{ying2018hierarchical} propose a differentiable graph pooling module to learn hierarchical representations of graphs on task of graph classification.

\section{Conclusion remarks}
In this paper, we study the problem of tree structure-aware graph representation learning. To effectively aggregate the multi-hop neighborhood, which usually present the structure of hierarchical trees in most heterogeneous graphs, we propose a tree structure-aware graph neural network model, which contains aggregation modules and sequence propagation modules to transform information of neighbors within multi-hop links to the root nodes. To overcome the problem of the mutual interference on the distributions of different types of nodes, we propose a relational metric learning module to optimize model, which utilize metrics to calculate node similarities on relation-specific metric spaces and embed nodes into type-specific spaces. From the experiment, the heterogeneous graph representation learning method we proposed can better extract the rich information of neighborhoods and improve the qualities of node representations.

\bibliographystyle{IEEEtran}
\bibliography{IEEEabrv,bibfile}
\end{document}